\documentclass[twocolumn]{aastex62}

\pdfoutput=1
\usepackage{natbib}
\usepackage{amsmath,amsfonts,amssymb,mathrsfs,graphicx,color}
\usepackage{verbatim}
\usepackage{enumerate}
\usepackage{graphicx}
\usepackage{xcolor}
\usepackage{slashed}
\usepackage[utf8]{inputenc}
\usepackage{float}

\graphicspath{ 
    {./figures_pdf/}    
}

\newcommand{\ie}{{\it i.e.}~}
\newcommand{\eg}{{\it e.g.}}
\newcommand{\eq}{Eq.~}
\newcommand{\Fig}{Fig.~}
\newcommand{\Sec}{Section~}
\newcommand{\Tab}{Tab.~}
\newcommand{\equ}[1]{\eq(\ref{eq:#1})}

\newcommand{\prince}{{\sc PriNCe}}
\newcommand{\xmax}[0]{X_\text{max}}
\newcommand{\deriv}[2]{\frac{\text{d} #1}{\text{d} #2}}

\submitjournal{\apj}

\begin{document}

\title{A new view on Auger data and cosmogenic neutrinos in light of different nuclear disintegration and air-shower models}

\author{Jonas Heinze}
\affiliation{DESY, Platanenallee 6, 15738 Zeuthen, Germany}

\author{Anatoli Fedynitch}
\affiliation{DESY, Platanenallee 6, 15738 Zeuthen, Germany}

\author{Denise Boncioli}
\altaffiliation{now at GSSI, viale Francesco Crispi, 67100, L'Aquila, Italy}
\affiliation{DESY, Platanenallee 6, 15738 Zeuthen, Germany}

\author{Walter Winter}
\affiliation{DESY, Platanenallee 6, 15738 Zeuthen, Germany}

\begin{abstract}

We study the implications of Ultra-High Energy Cosmic Ray (UHECR) data from the Pierre Auger Observatory for potential accelerator candidates and cosmogenic neutrino fluxes for different combinations of nuclear disintegration and air-shower models. We exploit the most recent spectral and mass composition data (2017) with a new, computationally efficient simulation code \prince{}. We extend a systematic framework, which has been previously applied in a combined fit by the Pierre Auger Collaboration, with the cosmological source evolution as an additional free parameter. In this framework, an ensemble of generalized UHECR accelerators is characterized by a universal spectral index (equal for all injection species), a maximal rigidity, and the normalizations for five nuclear element groups. We find that the 2017 data favor a small but constrained contribution of heavy elements (iron) at the source. We demonstrate that the results moderately depend on the nuclear disintegration (PSB, \textsc{Peanut}, or \textsc{Talys}) model, and more strongly  on the air-shower (\textsc{EPOS-LHC}, \textsc{Sibyll-2.3}, or \textsc{QGSjet-II-04}) model. Variations of these models result in different source evolutions and spectral indices, limiting the interpretation in terms of a particular class of cosmic accelerators. Better constrained parameters include the maximal rigidity and the mass composition at the source. Hence, the cosmogenic neutrino flux can be robustly predicted. Depending on the source evolution at high redshifts the flux is likely out of reach of future neutrino observatories in most cases, and a minimal cosmogenic neutrino flux cannot be claimed from data without assuming a cosmological distribution of the sources.
\end{abstract}


\section{Introduction}
\label{sec:intro}

The two largest detectors ever built, the Pierre Auger Observatory \citep{ThePierreAuger:2015rma} and the Telescope Array \citep{AbuZayyad:2012kk}, investigate the origin and the nature of Ultra-High Energy Cosmic Rays (UHECRs) above $10^{18}$ eV with hybrid detection techniques that combine signals from surface and fluorescence detectors to reconstruct extensive air showers, which are giant particle cascades initiated through interactions of the UHECRs with the atmosphere. There is evidence for an extragalactic origin of the UHECRs \citep{Aab:2017tyv}, and studies of the UHECR arrival directions uncovered interesting patterns such as a strong dipole anisotropy and a correlation with nearby source directions \citep{Aab:2018chp}. However, an association with a concrete source or class of sources is not yet in reach. The chemical composition is likely to be a mixture of different nuclear masses \citep{Aab:2016htd}, ranging from protons up to nitrogen or heavier nuclei \citep{Aab:2017cgk}. While the mass-sensitive experimental observables are statistically in agreement between the two experiments, their interpretation in terms of physical mass composition is still subject to discussions \citep{deSouza:2017wgx}.

Various astrophysical phenomena, typically associated with the emission of high-energy photons, have been proposed as potential accelerators of UHECRs. Gamma-Ray Bursts (GRBs), provided that a significant fraction of baryons is accelerated in their jets, can be capable of emitting UHECRs and producing also high-energy neutrinos due to photo-hadronic interactions of protons or heavier nuclei with the target photons \citep{Waxman:1997ti}. Blazars, a subset of powerful active galactic nuclei with their jets pointing at the observer, are numerous and powerful enough to sustain the UHECR spectrum and have been considered as sources of UHECRs and high-energy neutrinos \citep{Stecker:1991vm,Murase:2014foa,Rodrigues:2017fmu}. The absence of an associated neutrino signal in the IceCube detector \citep{Aartsen:2017wea,Aartsen:2016lir} constrains the density of cosmic rays in GRBs and blazars but does not necessarily exclude these classes of sources as UHECR accelerators. Other compact source classes, such as jetted Tidal Disruption Events (TDEs) \citep{Farrar:2014yla} or low-luminosity GRBs (LL-GRBs) \citep{Murase:2006mm}, are potentially luminous or copious enough to power the UHECR and high-energy neutrino sky. Starburst galaxies constitute a sample of sources in which the re-acceleration of PeV cosmic rays to ultra-high energies may occur at the termination shocks of kpc-scale ``super winds'' \citep{Anchordoqui:1999cu}. A higher abundance of young pulsars \citep{Blasi:2000xm} as an effect of an enhanced supernova rate might also predestine these galaxies as hosts of UHECR accelerators. The anisotropy observed by the Pierre Auger Observatory indeed indicates a directional correlation with a subset of nearby gamma-ray-bright starburst galaxies \citep{Aab:2018chp}. In all cases, the direct association with high-energy neutrinos would be a smoking gun signature for the origin of the cosmic rays. If, on the other hand, the neutrino production in the sources is inefficient, a directly related neutrino signal will be absent, and indirect methods will be needed to infer the nature of the cosmic ray accelerators. Obtaining information on the distribution of sources (such as their evolution as a function of redshift) is one such indirect method to identify the accelerators, and will be  therefore one of the main targets of our study.

The identification of the UHECR sources is complicated by the transport through the intergalactic medium (IGM) where interactions with the Cosmic Microwave Background (CMB) and Cosmic Infrared Background (CIB) photons alter the spectrum and chemical composition compared to the original emission at the source. 
By assuming a model for the UHECR spectra emitted from the sources and the extragalactic propagation through the IGM, one can infer the free source model parameters through a fit to the available UHECR data. In several such studies \citep{Hooper:2009fd,Aloisio:2013hya,Taylor:2015rla,Globus:2015xga,Aab:2016zth,Auger2017magfield} it has been assumed that the sources are identical, isotropically distributed and the UHECR emission follows power-law spectra with a rigidity-dependent cutoff.
Since these sources are representing {\it generic accelerators}, the cosmological evolution of the source density is undefined and requires one or multiple additional free parameters. Typically one assumes piece-wise defined evolution functions of the form $(1+z)^m$, with $m$ the evolution parameter. Due to accumulation of energy losses over large distances, UHECRs, even without considering magnetic fields, experience a {\it horizon} or maximal distance they can travel through the IGM, which is approximately equivalent to a redshift of $z \sim 1$, or a few Gpc. Therefore, the UHECR spectrum is almost insensitive to the parameterization of the source evolution beyond redshift $z \sim 1$. Interactions of UHECRs leave traces, namely cosmogenic neutrinos that are produced in photo-hadronic interactions with the target photons. Since neutrinos travel unimpeded through the IGM, the density of UHECRs for $z>1$ has an impact on their flux. As a consequence, the cosmogenic neutrino flux can be used to constrain the cosmological source evolution \citep{Ahlers:2009rf,Gelmini:2011kg,Aloisio:2015ega,Heinze:2015hhp,Romero-Wolf:2017xqe,AlvesBatista:2018zui,Moller:2018isk,Das:2018ymz,Wittkowski:2018giy,vanVliet:2019nse}.
 
The modeling of the transport comes with a number of uncertainties: photo-nuclear (photo-disintegration) reactions \citep{Batista:2015mea,Boncioli:2016lkt,Soriano:2018lly} that change the mass composition of nuclei due to interactions with CMB or CIB photons; the hadronic interactions, which are used in the interpretation of air-shower observables in terms of the mass composition; and the CIB spectrum, that is not well known at high redshifts. The interpretation of the UHECR data is affected by these uncertainties, as demonstrated in \citet{Batista:2015mea} and in the Combined Fit (CF) of the spectrum and composition data by the Pierre Auger Collaboration \citep{Aab:2016zth}. 

While in the CF different assumptions for source density evolutions have been tested for compatibility, no conclusions have been drawn about possible association with sources. Hence, the main attention was devoted to a flat cosmological evolution (non evolving source densities) \citep{Aab:2016zth}, which however cannot be easily related to known accelerator candidates. As an example, sources can evolve similarly to the star forming rate (SFR), $(1+z)^{3.4}$, for $z < 1$, such as GRBs \citep{2010MNRAS.406.1944W}.
Blazars have typically a more complicated luminosity-dependent evolution function and can evolve more steeply with redshift. Some source classes, such as TDEs, may have negative source evolutions.
As a consequence, any attempt to seek an astrophysical interpretation within the framework of such a fit requires the source evolution to be a free parameter. However, each new parameter is computationally expensive, which has led to different strategies to deal with this problem; for example, the redshift evolution can be included in a coarser way \citep{AlvesBatista:2018zui} or in a limited range of values \citep{Romero-Wolf:2017xqe} (see also \citet{Moller:2018isk,Das:2018ymz} for similar studies).

In this paper, we revisit the approach of the CF, taking into account the dominant model dependencies, and focus on the degeneracies between the fit parameters given a homogeneous distribution of generic UHECR sources. We study the impact of the model uncertainties on the astrophysical interpretation by performing scans in the three parameters: maximum rigidity $R_{\rm max}$ $[\mathrm{GV}]$ (corresponding to the maximum energy of acceleration divided by the charge of the particle, $E_{\rm max}/Z$), spectral index $\gamma$ and cosmological evolution index $m$, using different combinations of nuclear disintegration and air-shower models. The computational requirements are significantly reduced through the new numerical code \prince{}, \textit{Propagation including Nuclear Cascade equations}, that performs the propagation very efficiently under changing physical conditions. We are, therefore, able to investigate the full three-dimensional source parameter space with a comparable resolution in all parameters for different nuclear disintegration models. With Monte-Carlo or slower numerical codes such a study is not feasible due to excessive requirements of computational resources, and thus our result is novel. 
As an important result, we obtain the allowed parameter space contours that represent the state-of-the-art of current UHECR observations. Under the assumption of one dominant source population that accelerates cosmic ray nuclei up to a maximal rigidity, we accurately compute the expected cosmogenic neutrino fluxes and discuss the robustness of the predictions by studying the major model uncertainties.

\section{Models of UHECR transport and their sources}

In this section we describe the main model uncertainties affecting our analysis: The photo-backgrounds and cross-sections for the interactions during propagation, the hadronic interaction models used to infer UHECR properties from the observed air showers and the implied assumptions about the distribution and characteristics of UHECR sources.

\subsection{Extragalactic propagation}
\label{sec:disintegration_model}

During extragalactic propagation, UHECRs interact with the CMB and the CIB via photo-pair (e$^{+}$e$^{-}$) production and photo-nuclear processes. Additionally, all relativistic particles lose energy adiabatically due to the expansion of the Universe.
Photo-nuclear interactions can be subdivided into two regimes: photo-disintegration ($\varepsilon_r < 150$ MeV) and photo-meson production (above the pion production threshold, $\varepsilon_r > 150$ MeV), where $\varepsilon_r$ is the photon energy in the nuclear rest frame. 

In the photo-disintegration regime, the target photons interact with one or two nucleons and collectively excite the nucleus into a resonant state, which subsequently decays emitting (evaporating) nucleons, heavier fragments or keV-MeV photons. To model the cascading of secondary nuclei during propagation, numerical codes, such as \prince{}, described in \Sec\ref{sec:methods}, or Monte-Carlo packages, require as input inelastic interaction cross sections and inclusive cross sections (or multiplicities) of secondary particles. Such cross sections can be obtained either empirically from data as in the Puget-Stecker-Bredekamp (\textsc{PSB}) \citep{Puget:1976nz} parameterization, or by tabulating the output of more realistic nuclear models. In this study, we use \textsc{Talys} \citep{Koning:2007}, a comprehensive pre-equilibrium and Hauser-Feshbach theory based code, and \textsc{Peanut} \citep{FLUKA_pd1,FLUKA_pd2} -- an event generator of the \textsc{FLUKA} package \citep{Ferrari:2005zk,BOHLEN2014211} with an intra-nuclear cascade model at energies $\varepsilon_r > 200$ MeV and a similar set of statistical models below that (see \citet{Boncioli:2016lkt} for a discussion of these models and their uncertainties).

Qualitatively the distributions of secondaries are similar for the two statistical models, while quantitatively the results may vary depending on the availability of data for each individual isotope and the degree of parameter optimization for each of these isotopes. We observe that in the default configuration, \textsc{Peanut} is better optimized to the available data. Unofficial tables for \textsc{Talys} are available that can improve the description for some isotopes \citep{Batista:2015mea}. Compared to the \textsc{PSB} parameterization, where only one isotope for each mass number is used, \textsc{Peanut} and \textsc{Talys} demonstrate a faster disintegration into lighter elements, including the presence of heavier fragments (D, T, ${}^{3}$He, ${}^{4}$He). Therefore the interpretation of the UHECR data in terms of composition at the source is expected to vary with respect to the use of different disintegration models. 

Pion production off nuclei in all current propagation codes is handled in a ``superposition'' approach, \ie the nucleons are treated as quasi-free. The interaction cross sections and the pion yield for $\varepsilon_r > 150$ MeV scale as
\begin{equation}
    \sigma_{\rm A\gamma}(\varepsilon_r) = Z \sigma_{\text{p}\gamma}(\varepsilon_r) + N \sigma_{\text{n}\gamma}(\varepsilon_r) 
\end{equation}
with the number of protons $Z$ and the number of neutrons $N$.
The dominant pion production process is the $\Delta$-resonance production in the $s$-channel, $p + \gamma \rightarrow \Delta^+ \rightarrow p/n + \pi^0 / \pi^+$. The pion takes about $20\%$ of the primary's energy and results in significant energy losses for the projectile. In absence of other processes, the cutoff in the UHECR spectrum at $E \approx 4 \cdot 10^{10}$ GeV could be attributed to this energy loss, as predicted in \citet{Zatsepin:1966jv,Greisen:1966jv} and is referred to as the Greisen-Zatsepin-Kuzmin (GZK) cutoff. In the case of nuclei, the $\Delta$-resonance threshold is shifted by $A$ to higher energies. Instead most interactions take place at the energies of the Giant Dipole Resonance (GDR) around $\varepsilon_r \sim 20$ MeV leading to a cutoff in the spectrum of UHECR nuclei at energies similar to the GZK cutoff.

As cosmogenic neutrinos are only produced in the photo-meson regime, the differences between free nucleons and nuclei are striking. The photo-disintegration threshold prevents nuclei reaching energies $> A \cdot 10^{10}$ GeV where photo-meson production sets in on CMB target photons. Instead, pions and cosmogenic neutrinos are produced by nuclei at energies below the cutoff $\sim 10^9$~GeV on the less abundant CIB target photons. There are two consequences; the neutrino flux is expected to peak at lower energies $\sim 10^8$~GeV and to be significantly lower compared to the protons-on-CMB case. The impact of CIB variations on UHECR propagation has been studied in \citet{Batista:2015mea,Aab:2016zth}. While the effect on UHECR spectra is small, it becomes sizable for cosmogenic neutrino fluxes (see  \eg{} \citet{Aloisio:2015ega}).

Extragalactic and galactic magnetic fields play an important role at the ankle, which is the change of the spectral index at $5\cdot 10^{9}$ GeV \citep{Auger2017spectrum}, and below. The curvature of UHECR trajectories effectively elongates the distance to the sources. At sufficiently low rigidities ($\lesssim 10^{18}$ V) the particles are increasingly trapped in the neighborhood of their accelerator. The quantitative impact has been studied for example in \citet{Mollerach:2013dza}. It results in a hardening of the individual spectra of nuclei at lower energies at Earth and thus can soften the spectral index required at the source. In this work we neglect the effect of the magnetic fields, assuming a purely ballistic treatment of UHECR transport, as for example in \citet{Allard:2005ha,Hooper:2006tn,Aloisio:2008pp,Aloisio:2010he,Domenico:2013daa,Kalashev:2014xna,Aloisio:2017iyh} or in the one-dimensional version of CRPropa \citep{Batista:2016yrx}. While the deduced mass composition and source density evolution will remain almost unaffected, the spectral index may shift to softer values compared to what we show \citep{Auger2017magfield}. 

\subsection{Air-shower model}
\label{sec:air_shower_model}
When cosmic ray nuclei enter the atmosphere, the inelastic interactions with air molecules create hadronically (mesons and baryons) and electromagnetically (e$^\pm$ and photons) interacting particles with smaller energies. This cascading proceeds until most of the initial energy is dissipated as light and long lived particles (see \eg{} \citet{Matthews:2005sd} for an instructive model). The observation of the light and the secondary particles from these so-called extensive air showers allows the reconstruction of several properties of the original particle, such as the energy, the direction and to some extent the mass composition (see \citet{Kampert:2012mx} for a review). At the Pierre Auger Observatory and the Telescope Array, the energy is measured calorimetrically through the integration of the total fluorescence light yield. The direction is inferred through stereoscopy in combination with timing-based measurements at ground. The nuclear mass of the UHECR is the most challenging property, since it can only be derived indirectly by comparing a large number of observations with model-dependent simulations. Hence the measurement of the composition is a statistical argument.
\begin{figure}
    \centering
    \includegraphics[width=1.\columnwidth]{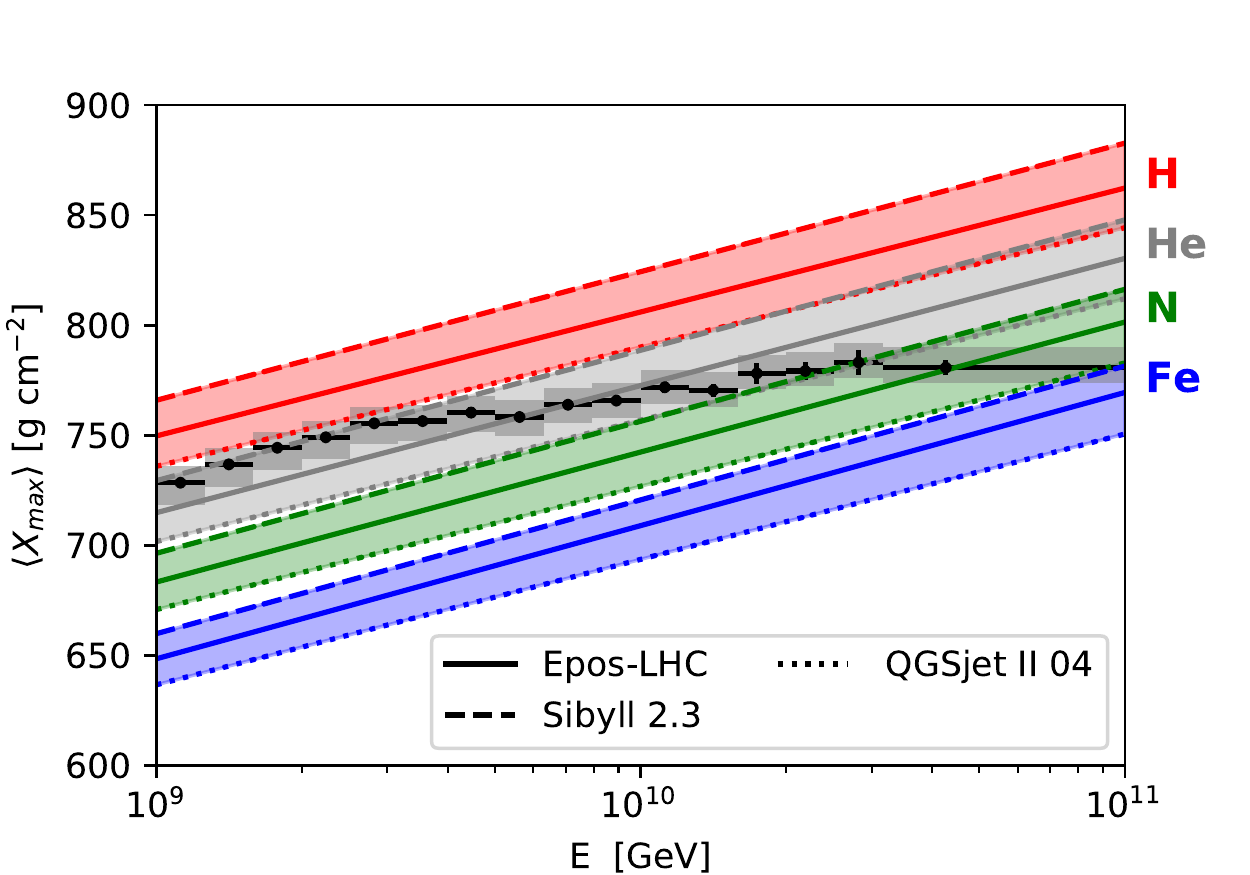}
    \includegraphics[width=1.\columnwidth]{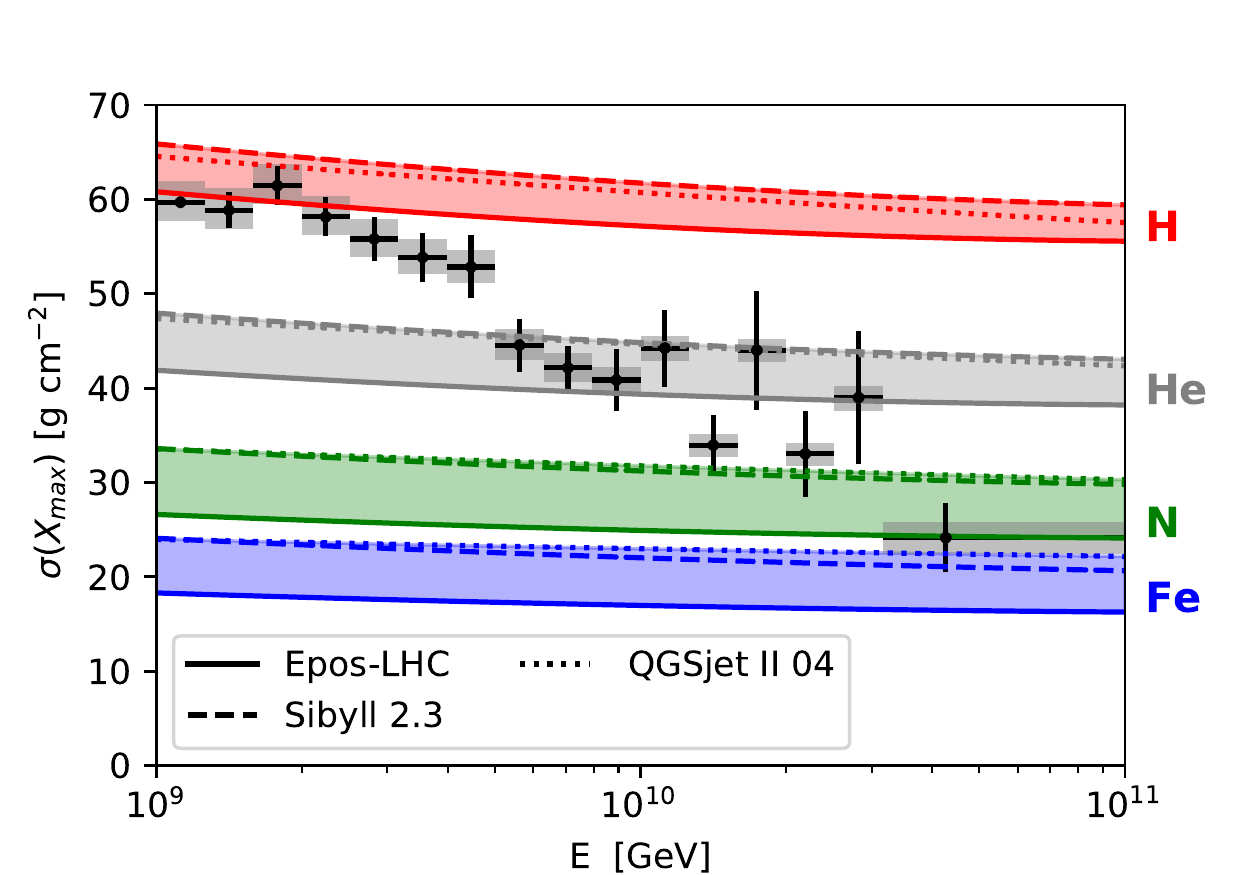}
    \caption{The Auger 2017 $\langle \xmax \rangle$ (top) and $\sigma(\xmax)$ data \citep{Auger2017xmax}, superimposed on different air-shower model expectations (\textsc{Epos-LHC} \citep{Pierog:2013ria}, \textsc{Sibyll-2.3} \citep{Riehn:2015oba} and \textsc{QGSjetII-04} \citep{Ostapchenko:2010vb}. The spread between the models (shaded areas) can be regarded as an interpretation uncertainty for the mass composition.}
    \label{fig:xmax_model_uncertainty}
\end{figure}

The sensitive variable for the mass composition is the $\xmax{}(E)$, the depth at which the energy dissipation of a single air shower is maximal. The $\xmax$ fluctuates, since the first interaction statistically occurs at different altitudes and because secondary particles can be produced with a multitude of multiplicity and energy configurations. The simplest description that captures the observed distributions is the combination of the mean $\langle \xmax \rangle$ and the dispersion or variance $\sigma(\xmax)$. The expected values are shown in \Fig\ref{fig:xmax_model_uncertainty} together with expectations for individual nuclei, obtained with different interaction models. 

Our simulations of the UHECR transport produce individual spectra for each nuclear mass at the top of the atmosphere for which we compute $\langle \ln{A} \rangle$ and $\sigma^2_{\ln{A}}$ at each energy of the numerical grid. We follow exactly the procedure from \citet[sec. 2]{Abreu:2013env} to convert the average of the logarithmic mass and its dispersion (bin-wise in energy) to the experimental observables $\langle \xmax \rangle$ and $\sigma(\xmax)$ using
\begin{equation}
    \label{eq:mean_xmax}
    \langle \xmax \rangle = \langle \xmax \rangle_\text{p} + f_E \langle \ln{A} \rangle,
\end{equation}
where $\langle \xmax \rangle_\text{p}$ is the mean depth at maximum of proton showers and $f_E$ parametrizes the dependence on the air-shower model and energy, and
\begin{equation}
    \label{eq:sigma_xmax}
    \sigma^2(\xmax) = \langle \sigma^2_\text{sh} \rangle + f_E^2 \sigma^2_{\ln{A}},
\end{equation}
where $\langle \sigma^2_\text{sh} \rangle$ contains the model- and $\langle \ln{A} \rangle$-dependent shower-to-shower fluctuations, while $\sigma^2_{\ln{A}}$ linearly depends on the dispersion of the masses; all parameters are dependent on the logarithm of the cosmic ray energy. The values of the parameters are obtained from air-shower simulations that do not take detector effects into account. Instead, this is taken into account by comparing with observables that are already corrected for detector effects. In contrast to the original paper \citep{Abreu:2013env}, we use an updated set of parameters for the post-LHC interaction models\footnote{Private communication with S. Petrera}.

Essentially, the first moment $\langle \xmax \rangle$ has a linear dependence on $\ln{A}$ where some non-linear effects are absorbed in $f_E$. When fitting the data, the different model expectations for $\langle \xmax \rangle_\text{p}$ impose shifts of the $\langle \ln{A} \rangle$ that are results of the propagation simulation and its initial conditions. The second term of the dispersion $\sigma^2(\xmax)$ becomes small if only a single mass is present, or, if spectra of similar/neighboring masses are superimposed. It is large in case a few masses with large distance in $\ln{A}$ dominate the sum of the spectra. The simultaneous description of both the mean and the variance of $\xmax$ is indispensable for any serious interpretation of the composition results since the variables are supplementary and sensitive to different features of the UHECR flux. 

For the present study, the differences in the conversion between mass and $\xmax$ observations are the most relevant feature of \Fig\ref{fig:xmax_model_uncertainty}. For example, at a fixed $\langle \xmax \rangle$, the $\langle \ln{A} \rangle$ inferred with \textsc{Sibyll 2.3} is heavier compared to the other models. At the same time the shower-to-shower fluctuations $\langle \sigma^2_\text{sh} \rangle$ in \equ{sigma_xmax} are high, implying strong constraints for the mass dispersion term $\sigma^2_{\ln{A}}$. While one can simply say ``\textsc{Sibyll 2.3} is heavier'' than \textsc{Epos-LHC}, the pulls on the fit induced by the properties of the models are highly non-trivial and discussed in a more ``applied'' way in section \ref{sec:model_dependence}. Note that some models, like \textsc{QGSJetII-04}, fail to produce a consistent relation between mass and $\xmax$ variables \citep{Aab:2014kda,Auger2017xmax}. 

\subsection{Source model}
\label{sec:source_model}

Several source candidates, in particular compact jetted sources, such as Gamma-Ray Bursts \citep{Globus:2014fka,Biehl:2017zlw,Zhang:2017moz,Boncioli:2018lrv} or Tidal Disruption Events~\citep{Zhang:2017hom,Guepin:2017abw,Biehl:2017hnb}, can describe the UHECR spectrum and composition. Another category of viable UHECR sources are starburst galaxies \citep{Anchordoqui:1999cu,ANCHORDOQUI2019} that may also contain populations of powerful accelerators \citep{Fang:2013cba}. The majority of models assume Fermi acceleration as the dominant acceleration process, yielding a power law with spectral indices close to $\gamma=2$ at the acceleration site. Hence, charged particles are magnetically confined at the site of acceleration leading to an additional modification of the spectrum due to the escape mechanism. For example, diffusive or direct escape harden the in-source flux by up to one power \citep{Baerwald:2013pu}, while advective escape may act as a low-pass filter and suppress the high-energy emission in the presence of a sizable cooling process \citep{Murase:2014foa}. More sophisticated simulations suggest even harder, bell-shaped, escape spectra \citep{Ohira:2009rd,Globus:2014fka}. Other acceleration mechanisms have been proposed that result in almost monochromatic particle spectra \citep{Lyubarsky:2000yn,Kirk:2017aon}. Therefore, spectra of escaping charged particles that are significantly harder than $E^{-2}$ are not unexpected for a single source. However, we note that in the current approach, we consider an entire ensemble of sources, and it seems unlikely that all sources will behave in the same way, \ie\ reach the same maximal rigidity and have the same mass composition. Therefore, too hard or even peaked ($\gamma < 0$) spectra may be difficult to reconcile with current knowledge.

In the interest of comparability we parameterize our generic source population exactly the same way as in the CF \citep{Aab:2016zth} and \citep{AlvesBatista:2018zui}:
\begin{align}
    \label{eq:injection}
    J_A(E) = \mathcal{J}_A~f_\text{cut}(E,Z_A,R_\text{max})~n_\text{evol}(z)\left(\frac{E}{10^9~\text{GeV}} \right)^{- \gamma},
\end{align}
in which the nuclear species $A$ (here ${}^{1}$H, ${}^{4}$He, ${}^{14}$N, ${}^{28}$Si and ${}^{56}$Fe) share a common spectral index $\gamma$ and a maximal rigidity $R_{\mathrm{max}} = E_\mathrm{max}/Z_\text{A}$. The $\mathcal{J}_A$ are free normalization constants representing the number of particles ejected from the sources per unit of time, comoving volume and energy. The functional form of the cutoff is arbitrary and we adopt the definition of the CF:
\begin{align}
    f_\text{cut}(E) = \begin{cases}
        1 
        &, E < Z_\text{A} R_\text{max} \\
        \exp\left(1 - \frac{E}{Z_\text{A} R_\text{max}} \right)
        &, E > Z_\text{A} R_\text{max}.
    \end{cases}
\end{align}

In the CF the fractions of injection elements $f_\text{A}$ are defined at a fixed energy point ($10^9$~GeV), relative to a total normalization. This definition is easily obtained from our $\mathcal{J}_A$ as $f_A = \mathcal{J}_A / \sum_A \mathcal{J}_A$.

A physically more meaningful definition of the mass fractions, that does not depend on the arbitrary choice $10^9$~GeV in \equ{injection}, is the integral fraction of the energy density
\begin{align}
    \label{eq:integral_fractions}   
    I^{9}_A = \frac{\int_{E_\text{min}}^\infty J_A(E) E \text{d}E}{\sum_A \int_{E_\text{min}}^\infty J_A(E) E \text{d}E} \, ,
\end{align}
where we choose $E_\text{min} = 10^{9}~\text{GeV}$ as the lower boundary. We will mostly refer to $I^{9}_A$, providing the $f_A$ for comparability with the CF.

In \equ{injection}, the parameterization for the source evolution with redshift is given by the function:
\begin{align}
    n_\text{evol}(z) = (1 + z)^{m} .
\end{align}
For variable $m$, the function approximates all known continuous source density functions within the UHECR horizon $z \lesssim 1$. However for the prediction of other messengers it needs to be extrapolated to higher redshifts. In connection with the cosmogenic neutrino estimates (see section \ref{sec:cosmo_neu}), we will adopt more complex source distributions that include breaks. 

This flexible parameterization catches many features of theoretical source spectra. However, one has to keep in mind that the assumption of a rigidity-dependent escape is relatively strong and applies only to a subset of sources in which the maximal energy is limited by the size of the source rather than by cooling processes \citep{Biehl:2017zlw,Rodrigues:2017fmu}. Another impacting assumption is that of single dominant source population.
The complexity can be increased by accounting for an additional proton component with higher rigidity \citep{vanVliet:2019nse} or even by a detailed modelling of individual nearby sources \citep{Eichmann:2017iyr}. This however also vastly increases the degrees of freedom, making a global fit of all free parameters unfeasible given the current statistics of the UHECR data. 

\section{Simulation methods}
\label{sec:methods}

In this section we decribe methods of our global fit: the method used for the calculation of UHECR propagation through the IGM and the global fit of the propagated spectra to the observed data.

\subsection{Propagation of UHECRs with \prince{}}
\label{sec:propagation}

To study the model dependencies in photo-nuclear cascades, we developed a new original computer code called \prince{} (\textit{Propagation including Nuclear Cascade equations}) to efficiently solve the cosmic ray transport problem. Instead of the Monte-Carlo methods used in public codes such as \textsc{CRPropa} \citep{Batista:2016yrx} or \textsc{SimProp} \citep{Aloisio:2017iyh}, \prince{} numerically solves a system of coupled partial differential equations (PDEs) for the comoving density $Y_i(E_i, z)$ for each particle species $i$
\begin{align}
    \begin{split}
        \label{eq:transport_eq}
        \partial_t Y_i = 
        & -\partial_E (b_\text{ad} Y_i) - \partial_E \left(b_{e^+ e^-}  Y_i \right)\\
        & - \Gamma_i Y_i + \sum_j Q_{j \rightarrow i} (Y_j) + J_i.
    \end{split}
\end{align}
for an arbitrary distribution $J_i(E,z,A_i)$ of isotropically emitting and homogeneously distributed cosmic ray sources. The terms (in order of occurrence) represent adiabatic cooling, pair production, photo-nuclear interactions (interaction and decays; reinjection) and injection from sources. The system of PDEs in $E$ and $z$ is solved using a 6$^{\rm th}$-order finite difference operator for the $E$ derivatives and backward differentiation functions (BDF), essentially an iterative implicit solver, for the redshift dependence.\footnote{{\sc scipy.integrate.BDF} \url{https://docs.scipy.org/doc/scipy/reference/generated/scipy.integrate.BDF.html}} The latter is required since \equ{transport_eq} becomes stiff in $z$ for nuclear systems (more details on the code and the numerical methods are given in Appendix \ref{sec:prince_code}).

\eq(\ref{eq:transport_eq}) is only valid under the assumption of a homogeneous source distribution with a separation much smaller than the diffusion length. For this case the diffusion in extragalactic magnetic fields can be neglected. This reduces the calculation exclusively to the ballistic regime, in which the propagation becomes a one dimensional problem (time or redshift). This approximation in particular makes sense if one is interested in the highest energies above the ankle, where the impact of diffusion is small.

While similar codes have been previously developed, as for example in \citet{Allard:2005ha,Hooper:2006tn,Aloisio:2008pp,Aloisio:2010he,Domenico:2013daa,Kalashev:2014xna}, our code stands out due to its very high computational speed and numerical precision. Even without significant architectural optimizations, \prince{} performs the computation of nuclear and neutrino spectra within 30 seconds on a single core, integrating an arbitrary injection spectrum that can contain elements with $A \leq 56$ from a redshift of $z=1$. While Monte-Carlo techniques for UHECR propagation become efficient due to the possibility of re-weighting of pre-computed events, our code shines when interest is devoted to model uncertainties, since we can essentially change any parameter and recompute within these 30s, taking into account the impact on all relevant interaction rates. This includes arbitrary variations of the target photon densities without relying on simplified redshift scaling assumptions as often employed in Monte-Carlo methods or common numerical approaches. A detailed description of the numerical methods in \prince{} can be found in Appendix \ref{sec:prince_code}.

\subsection{Simulation and fitting procedure}
\label{sec:simulation}

This section aims to summarize the relevant setup of the simulations. We choose the five representative injection elements: hydrogen ($^1$H), helium ($^4$He), nitrogen ($^{14}$N), silicon ($^{28}$Si) and iron ($^{56}$Fe) in accordance with the CF. We verified that choosing different injection elements of the same mass groups yields qualitatively similar results. The generic source model has eight free parameters: $R_{\rm max}$, $\gamma$, $m$ and free normalizations $\mathcal{J}_A$ corresponding to the five injection elements. We allow for a shift $\delta_\text{E}$ in energy within the systematic uncertainty given by Auger ($\pm 14 \%$) \citep{Auger2017spectrum}.

The transport equation (\equ{transport_eq}) is linear in the normalization factor $\mathcal{J}_A$ but not in the other source parameters ($\gamma$, $R_\text{max}$ and $m$), triggering us to employ a two-staged approach for the fit.

In the first stage, we discretize the parameter space for $\gamma$, $R_{\rm max}$ and $m$ with these ranges and granularity:

\begin{center}
\begin{tabular}{l | r | r | r }
\hline \hline
& min & max & stepsize \\ \hline 
$\gamma$       & -1.5 & 2.5 & 0.05 \\
$\log_{10}(R_\text{max})$ & 9.7 & 11.7 & 0.05 \\
$m$            & -6 & 6 & 0.2 \\
\hline
\end{tabular}
\end{center}

For each point of this three-dimensional (3D) source-parameter grid, we separately compute the spectra at Earth for the five injection elements ($\sim1.5 \cdot 10^6$ individual simulations for one choice of the photo-nuclear interaction model). Since the propagated spectra are linear in the $\mathcal{J}_A$'s, the all-particle spectrum is calculated as a linear superpostion of the results obtained for single element injection.

In the second stage, we fit the nuclear fractions $\mathcal{J}_A$ and energy shift $\delta_\text{E}$ to the spectrum and the first two moments of $\xmax$ for each triplet in $(R_{\rm max}, \gamma, m)$ using the \textsc{Minuit} package\footnote{We use the \textsc{iMinuit} interface \url{https://github.com/iminuit/iminuit}.} \citep{1975CoPhC..10..343J}.
The translation from individual mass spectra at the top of the atmosphere to $\langle X_\text{max} \rangle$ and $\sigma(X_\text{max})$ is performed with the parameterization from \citet{Abreu:2013env}, using updated parameter sets for \textsc{Sibyll 2.3} and \textsc{Epos-LHC}.

To find the $\chi^2$ values for the UHECR fits within the entire 3D parameter space, the simulations are performed starting from redshift $z_{\mathrm{max}} = 1$. Once the $3\sigma$ confidence intervals are localized, we run additional simulations starting from $z=3$ to compute cosmogenic neutrino fluxes, verifying that the previously derived contours are unaffected by higher redshifts. Both stages have to be repeated for each propagation model, while a change of the air-shower model only requires the repetition of the second stage. In all cases, the CIB model is fixed to Gilmore et al. \citep{Gilmore:2011ks}.

The following $\chi^2$ definition is used as the goodness of fit estimator:
\begin{align}
    \chi_\mathcal{F}^2 &= \sum_i \frac{(\mathcal{F}(E_i) - \mathcal{F}_\text{model}(E_i,\delta_E))^2}{\sigma_i^2},
\end{align}
where $\chi_\mathcal{F}^2$ refers to each of the three observables $\mathcal{F}$, namely the combined spectrum, $\langle X_{\mathrm{max}}\rangle$ and $\sigma(X_{\mathrm{max}})$. The total $\chi^2$ is obtained by summing. A nuisance parameter $\delta_E$ captures the uncertainty in the energy scale, and we assume its distribution to be flat within $\pm 14\%$. The fit takes into account all data points above ${E_\text{min} = 6 \cdot 10^9 ~ \text{GeV}}$. The global best fit $\chi^2_\text{min}$ is found by minimizing over all points of the 3D parameter space.

We then use $\Delta \chi^2 = \chi^2 - \chi^2_\text{min}$ to draw contours around the best fit point by projecting to planes of two parameters by minimizing over all other parameters of the scan. While this frequentist approach is sufficient to draw contours and discuss the correlations among source parameters, there are more physical model parameters originating from the combination of discrete model choices, such as that for the photon background, the disintegration and the hadronic interaction model. 
We did not attempt to parametrize these model uncertainties by continuous nuisance parameters, as these are impossible to define in a physically meaningful and unbiased sense. We therefore choose discrete model combinations and discuss their qualitative differences in the fit contours.

\section{Impact of the updated 2017 data set on the two-dimensional fit}
\label{sec:flat_evolution_fit}

\begin{figure*}
    \centering
    \includegraphics[width=\columnwidth]{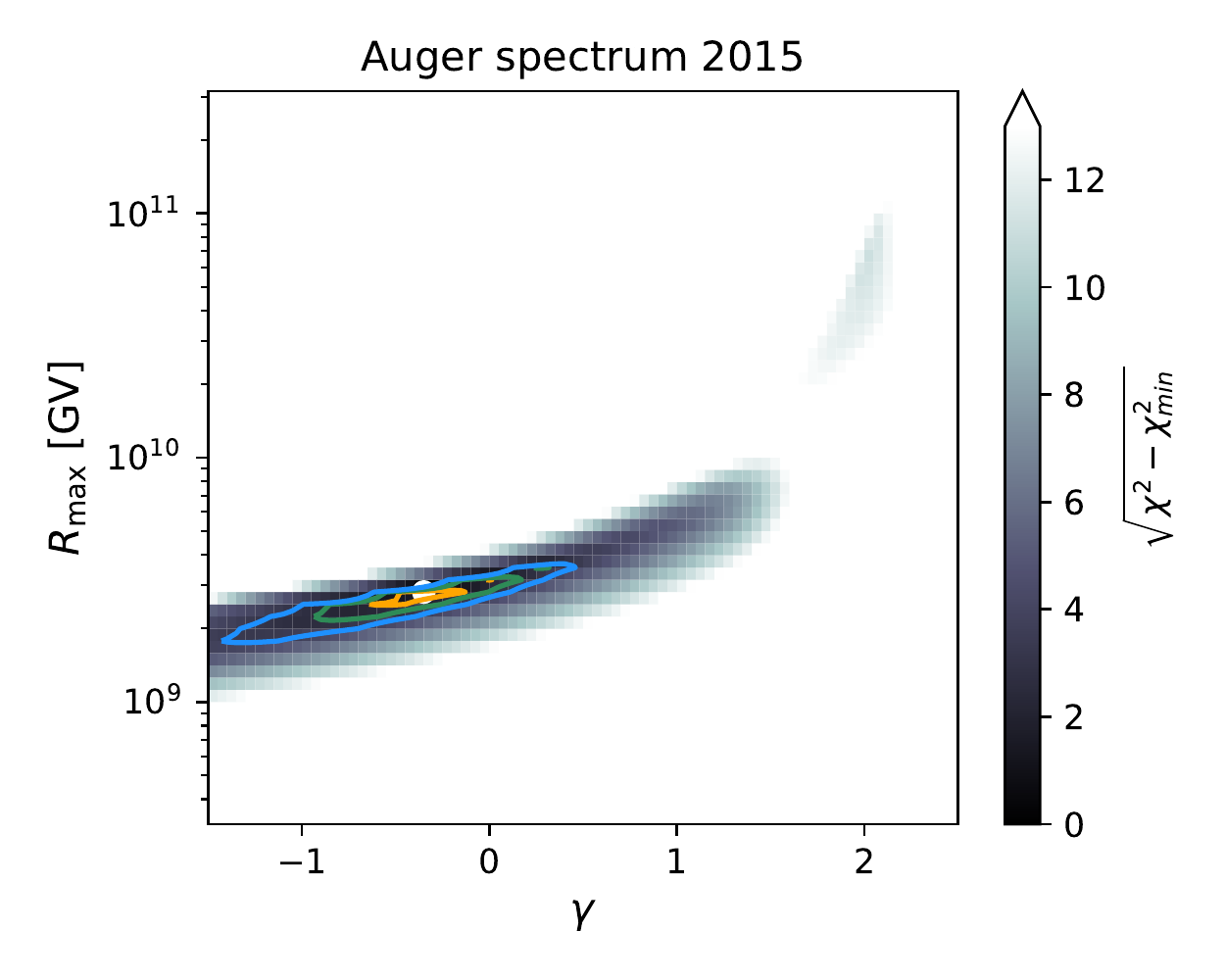} 
    \includegraphics[width=\columnwidth]{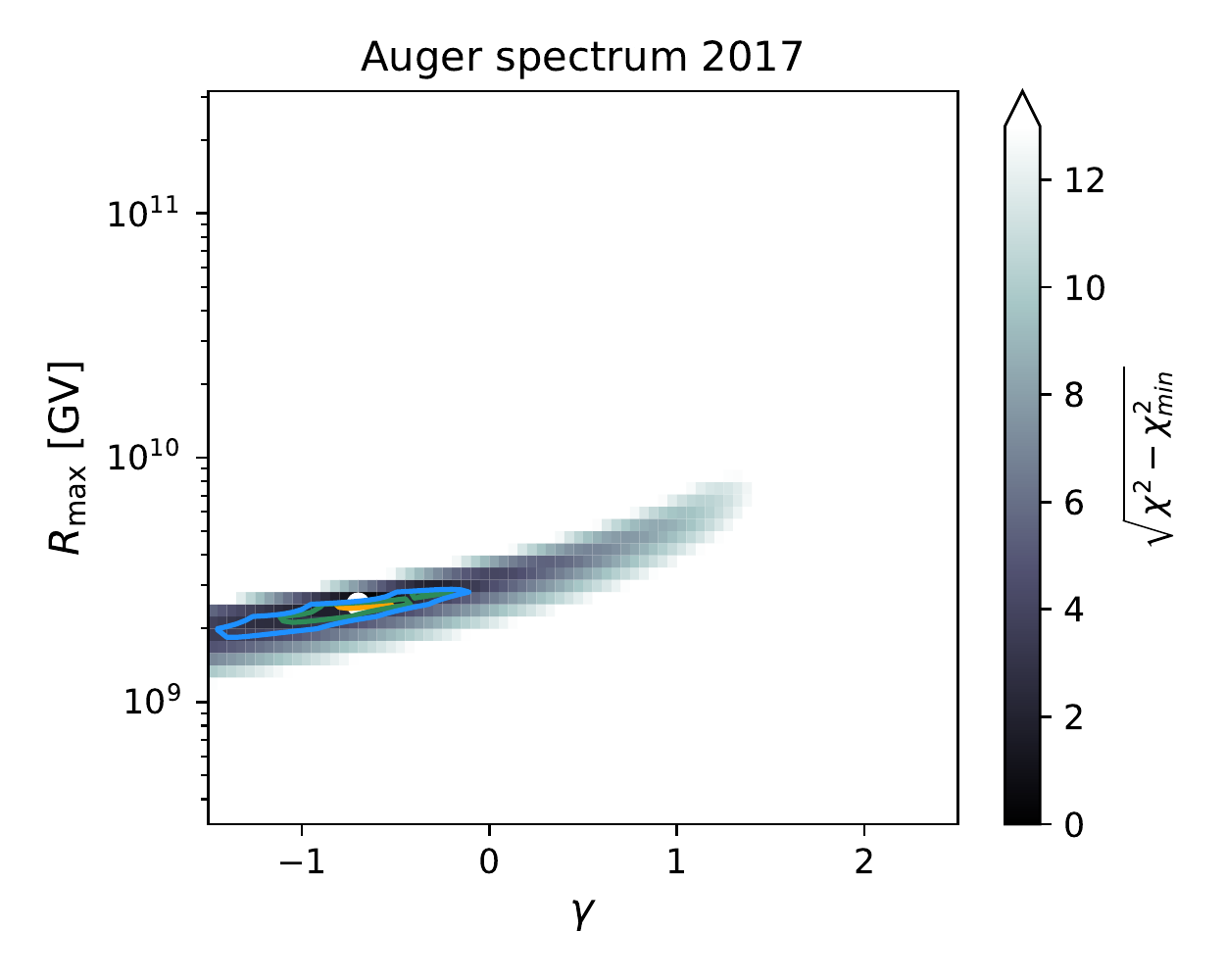}
    \caption{Parameter space in $\gamma$ and $R_{\mathrm{max}}$ assuming flat source evolution, corresponding to the Auger data sets of 2015 (left) and 2017 (right). The fits were done for model combination of \textsc{PSB} and \textsc{Epos-LHC} and with fixed energy scale, to be directly comparable to the CF baseline case \citep{Aab:2016zth}. The colored shading corresponds to $\sqrt{\chi^2 - \chi^2_\text{min}}$, while ${\chi^2 - \chi^2_\text{min}}$ is used to determine the contours. The contours are given for $1\sigma, 2\sigma, 3\sigma$ (for 2 d.o.f.). The best fit in each panel is indicated by a white dot.}
    \label{fig:paraspace_flat}
\end{figure*}

\begin{figure*}
    \centering
        \includegraphics[width=.45\textwidth]{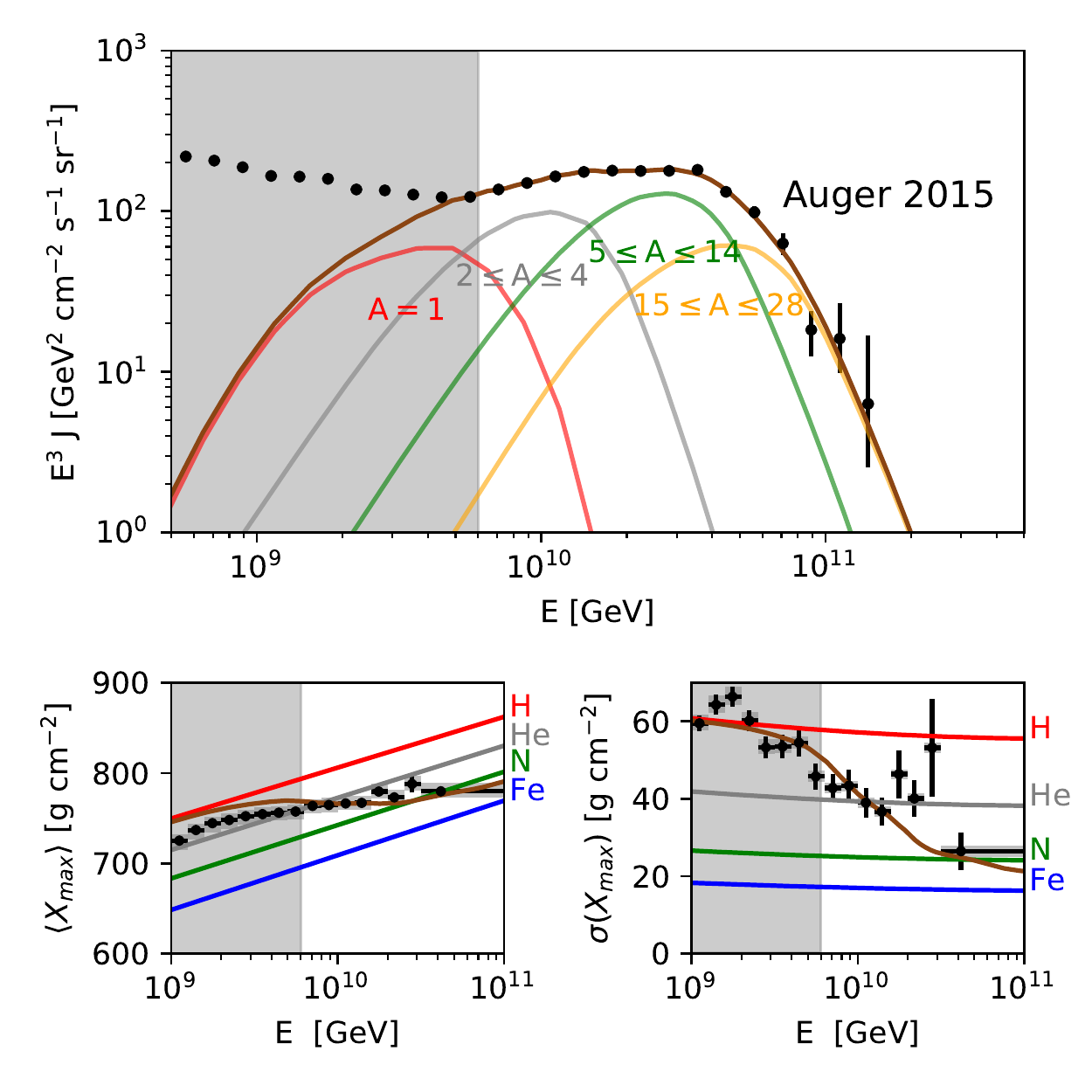}
        \includegraphics[width=.45\textwidth]{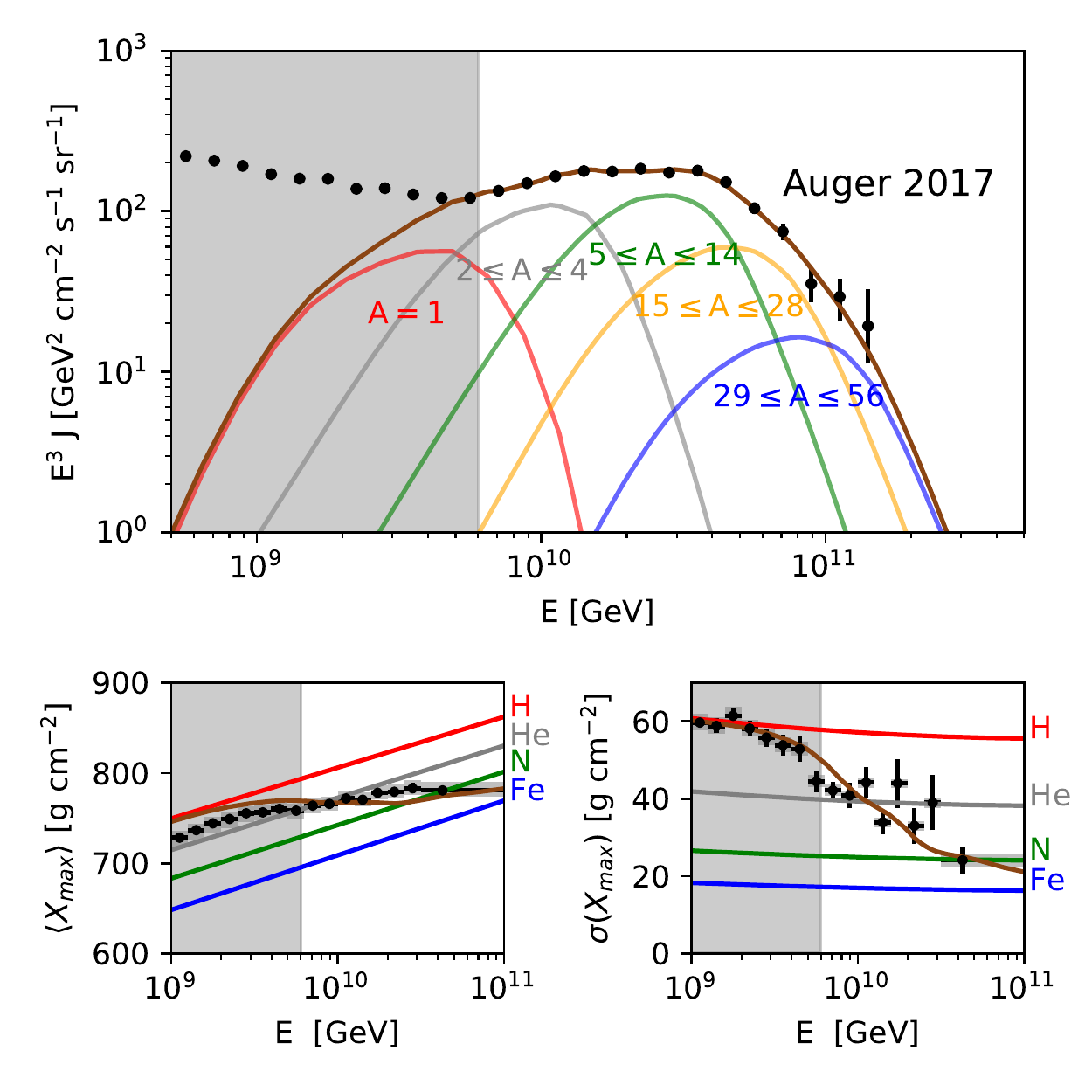}
    \caption{Spectra (upper panels) and composition observables (lower panels) corresponding to the best fit to the Auger 2015 (left) and 2017 (right) data assuming flat source evolution and scanning in $R_\text{max}$ and $\gamma$. The best fit values are found at $\gamma = -0.35$, $R_\text{max} = 2.8 \cdot 10^{9}$~GV (2015 data) and $\gamma = -0.7$, $R_\text{max} = 2.5 \cdot 10^{9}$~GV (2017 data). The gray shaded area indicates the range below $6 \cdot 10^9$ GeV, which is excluded from the fit. The expected composition is calculated assuming the EPOS-LHC shower model and comparing to the first two moments of $X_\text{max}$ distributions.}
    \label{fig:bestfit_flat}
\end{figure*}

\begin{deluxetable}{l|cc|cc}
    \tablecaption{Best fit parameters corresponding to the results of the fit with flat source evolution for the combination of \textsc{PSB} and \textsc{Epos-LHC}, using the 2015 and 2017 Auger data sets. For $\gamma$ the $1\sigma$-uncertainty (for 1 d.o.f.) is given. No uncertainty on $R_\text{max}$ is reported, as our computation grid is too sparse to resolve it.}
    \label{tab:bestfits_old_new_data}
    \tablehead{
        &\twocolhead{Auger 2015}
        &\twocolhead{Auger 2017}
    }
    \startdata
    $\gamma$
        & \multicolumn2{r|}{$-0.35_{-0.08}^{+0.15}$}
        & \multicolumn2{r}{$-0.70_{-0.08}^{+0.12}$}
    \\
    $R_\text{max}$ (GV)
        & \multicolumn2{r|}{$(2.8 \pm 0.2) \cdot 10^{9}$}
        & \multicolumn2{r}{$(2.5 \pm 0.1) \cdot 10^{9}$}
    \\
    $m$
        & \multicolumn2{r|}{$0.0$ (fixed)}
        & \multicolumn2{r}{$0.0$ (fixed)}
    \\
    $\delta_E$
        & \multicolumn2{r|}{$0.0$ (fixed)}
        & \multicolumn2{r}{$0.0$ (fixed)}
    \\
    \hline
    $f_A (\%)$
        & H  & He & H  & He
        \\
        & $5.8_{-5.8}^{+22.0}$
        & $89.9_{-0.7}^{+0.6}$
        & $9.7_{-9.7}^{+17.1}$
        & $87.8_{-0.6}^{+0.5}$
        \\
        & N & Si & N & Si
        \\
        & $4.0 \pm 0.2$
        & $0.3 \pm 0.0$
        & $2.4 \pm 0.2$
        & $0.1 \pm 0.0$
        \\
        & \multicolumn{2}{c|}{Fe}
        & \multicolumn{2}{c}{Fe}
        \\
        & \multicolumn{2}{c|}{$0.0_{-0.0}^{+4.6} \cdot 10^{-3}$}
        & \multicolumn{2}{c}{$(3.7 \pm 2.0) \cdot 10^{-3}$}
    \\
    \hline
    $I^{9}_A (\%)$
        & H  & He & H  & He
        \\
        & $0.6_{-0.6}^{+3.0}$
        & $46.7_{-1.8}^{+1.6}$
        & $0.8_{-0.8}^{+1.9}$
        & $47.9_{-1.4}^{+1.3}$
        \\
        & N & Si & N & Si
        \\
        & $39.9_{-1.3}^{+1.2}$
        & $12.8_{-1.2}^{+1.1}$
        & $37.9_{-1.6}^{+1.5}$
        & $11.4_{-2.3}^{+2.2}$
        \\
        & Fe & & Fe
        \\
        & $0.0_{-0.0}^{+1.0}$
        &
        & $2.1 \pm 1.1$
    \\
    \hline
    $\chi^2$ / dof
        & \multicolumn{2}{r|}{44.4 / 22}
        & \multicolumn{2}{r}{65.3 / 22}
    \\
    \enddata
\end{deluxetable}

We start the discussion of our results from the state of the CF and study the interesting impact of the updated 2017 data set \citep{Auger2017spectrum,Auger2017xmax} by reproducing a similar procedure to the one in \citet{Aab:2016zth} with our new code \prince{}. The source evolution parameter is fixed to $m=0$ (flat evolution); the nuclear disintegration, the CIB and the air-shower model are fixed to \textsc{PSB}~\citep{1975CoPhC..10..343J}, Gilmore et. al.~\citep{Gilmore:2011ks} and \textsc{Epos-LHC}~\citep{Pierog:2013ria}, respectively. The minimization runs over the spectral index $\gamma$, $R_\text{max}$ and the nuclear fractions $\mathcal{J}_A$. The energy scale is fixed and not allowed to float within its systematic uncertainty.

The energy range of the CF starts at $5 \cdot 10^{9}$ GeV. We noticed that with the new data set, $\chi^2$ is significantly affected by the small discontinuity next to the $\langle X_\text{max} \rangle$ point at $5.5 \cdot 10^{9}$ GeV, \ie this point alone adds a $\chi^2 \approx 35$ to the best fit with a total $\chi^2 \approx 102$. We therefore treat this data point as an outlier and start our fit range at $6 \cdot 10^{9}$~GeV, which does otherwise not qualitatively impact the fit.

The contours are shown in \Fig\ref{fig:paraspace_flat} and the best fit values are summarized in \Tab\ref{tab:bestfits_old_new_data}.
For the 2015 data set we find the same qualitative result as the CF: a flat extended minimum with $\gamma < 1$ and ${1 \cdot 10^9 < R_\text{max} < 8 \cdot 10^9}$~GV, and a second local minimum at $\gamma \approx 2$ and ${R_\text{max} \approx 4 \cdot 10^{10}}$~GV.
The differences in the exact locations of the minima with respect to the CF can be explained by the different propagation code used, as already pointed out in \citet{Aab:2016zth}. Additional small shifts originate from the use of the experimental observables. While we fit the first two $X_\text{max}$ moments for the composition, the CF uses the full $X_\text{max}$ distribution. This has the strongest impact on the second minimum at $\gamma = 2$, which becomes less significant in our approach. In addition, we directly fit the combined unfolded spectrum and do not use a forward-folding procedure in the fit.

When switching to the 2017 data set, the best fit parameters do not qualitatively change (see \Tab\ref{tab:bestfits_old_new_data}). However, the $\chi^2$ becomes worse due to the higher statistics. The allowed contours become narrower with a stronger preference for positive spectral indices. The second local minimum disappears. The reasons are the reduced statistical errors and a narrower width of the $X_{\mathrm{max}}$ distribution at the highest energies of the 2017 data set, leaving less room for the combination of a high $R_\text{max}$ with somewhat softer spectral indices. 

The largest qualitative difference concerns the injected iron fraction. While the 2015 data set did not require iron at the source, the new data suggest a small - but non-zero - integral iron fraction $I^{9}_\text{Fe} \approx 2\%$.
This is also visible in the comparison of the best fit spectra in \Fig\ref{fig:bestfit_flat}: for the 2017 data set (right panel) there is a contribution of heavy elements at the cutoff, which is absent in the fit to the 2017 data set (left panel).
This is due to the higher statistics of the three highest energy data points in the spectrum, which lead to a hardening. Due to the low rigidity found in the fit, reaching these energies requires a high charge number and therefore a significant iron fraction. However, this relies on the assumption of the rigidity dependence of the maximal energy and the fixed energy scale and hence cannot be rigorously interpreted as evidence for a non-zero iron fraction. Note, however, it will be still visible if we  later let the energy scale float. An indication for an iron contribution might also be visible in the composition data above $10^{19.4}$ eV \citep{ObservatoryMichaelUngerforthePierreAuger:2017fhr}.

\section{Three-dimensional fit}

We now include the source evolution $m$ as an additional free parameter and allow the energy scale $\delta_E$ to float within the systematic uncertainties by following the procedure described in \Sec\ref{sec:methods}.
First we discuss our ``baseline'' case defined by the combination of \textsc{Talys} as disintegration model and \textsc{Sibyll~2.3} as air-shower model (in \Sec\ref{sec:baseline_fit}), before extending to other model combinations (in \Sec\ref{sec:model_dependence}).
The impact of the model choices on the injected composition is discussed in \Sec\ref{sec:injection_fractions}.
  
\subsection{Baseline case characteristics}
\label{sec:baseline_fit} 
\begin{figure*}
    \centering 
    \includegraphics[width=0.7\textwidth]{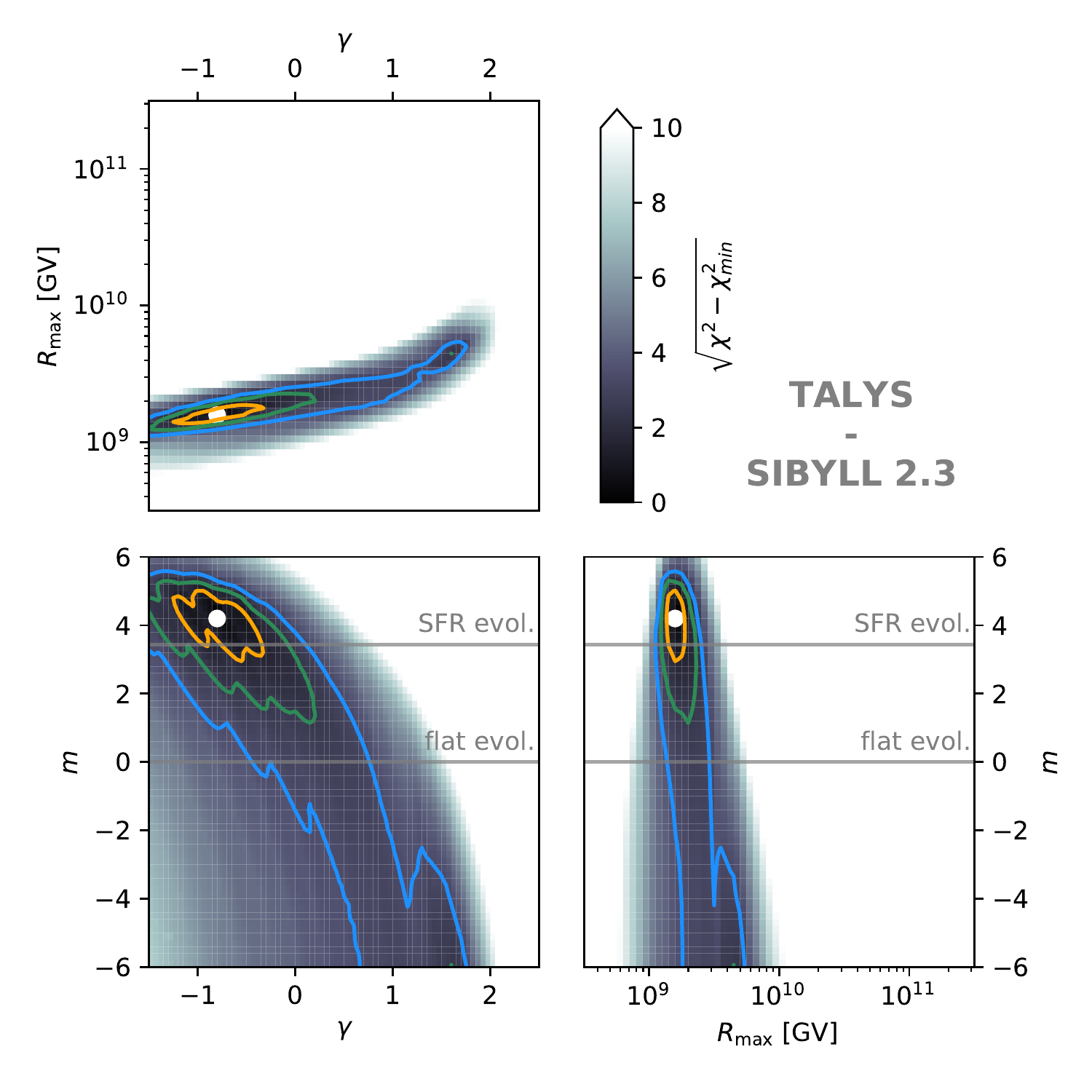}
    \caption{Parameter space in $\gamma$, $R_\text{max}$ and $m$ for the model combination \textsc{Talys} and \textsc{Sibyll 2.3} (baseline case) with free energy scale as a nuisance parameter. The best fit, found at $\gamma = -0.8$, $R_\text{max} = 1.6 \cdot 10^9$~GV and $m = 4.2$, is marked by a white dot. The colored shading corresponds to $\sqrt{\chi^2 - \chi^2_\text{min}}$, while ${\chi^2 - \chi^2_\text{min}}$ is used to determine the contours, which are given for $1 \sigma, 2 \sigma, 3 \sigma$ (for 2 d.o.f.). In each 2D panel, the third parameter is treated as a nuisance parameter and minimized over to project the 3D parameter space.}
    \label{fig:paramspace_3D}
\end{figure*}

\begin{figure}
    \centering
    \includegraphics[width=.49\textwidth]{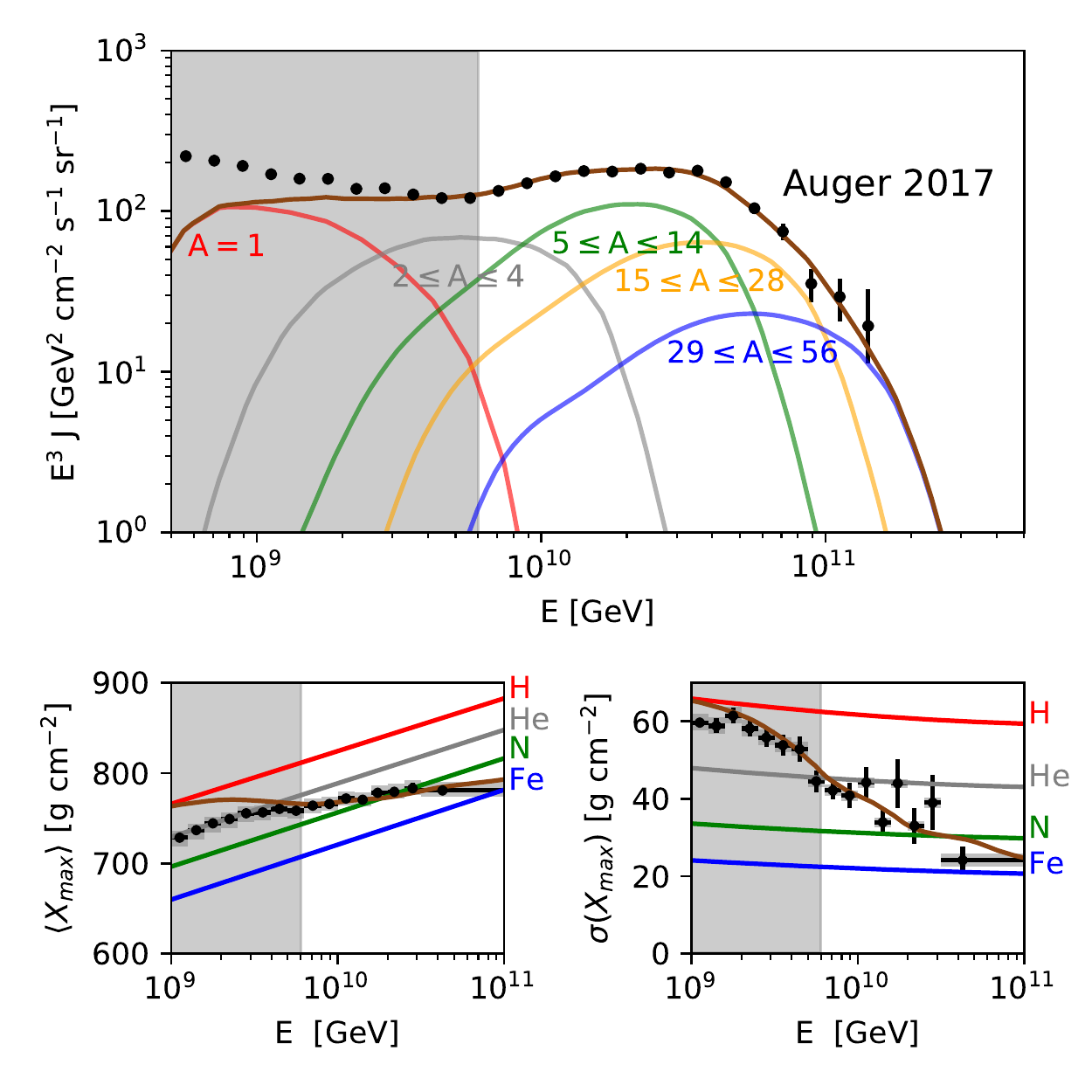}
    \caption{Spectrum (upper panel) and composition observables (lower panels) corresponding to the best fit to the Auger 2017 data, for the baseline model combination \textsc{Talys} and \textsc{Sibyll~2.3}. The corresponding injection at the source is found in \Fig\ref{fig:injection_bestfit}.}
    \label{fig:bestfit_3D}
\end{figure}

\begin{figure}
    \centering
    \includegraphics[width=\columnwidth]{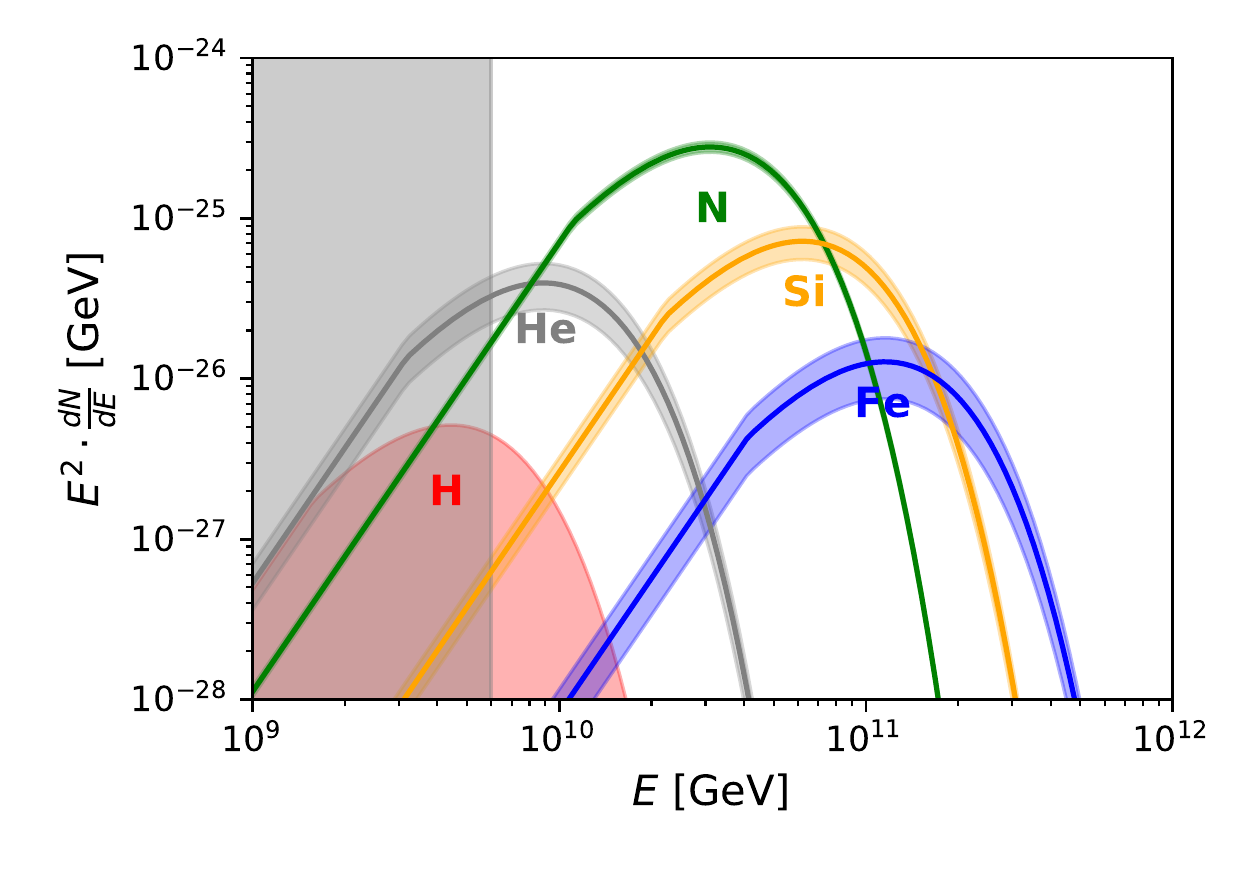}
    \caption{Injection spectra for the five injected elements corresponding to the best fit for the 3D parameter scan in \Fig\ref{fig:bestfit_3D} ($\gamma = -0.8$, $R_\text{max} = 1.6 \cdot 10^9$~GV, $m=4.2$). The shaded regions indicate the $1 \sigma$ uncertainties to the normalization of each injection corresponding to the fit (for $\gamma, R_\text{max}, m$ fixed). While the best fit proton fraction is 0, there can be a significant proton contribution within the uncertainty.}
    \label{fig:injection_bestfit}
\end{figure}

\begin{deluxetable}{l|ccc}[b]
    \tablecaption{\label{tab:bestfits3D} Best fit parameters for the 3D parameter scan with free source evolution for the baseline case of the combination \textsc{Talys} - \textsc{Sibyll~2.3}. For $\gamma$, $m$ and $\delta_E$ the $1\sigma$-uncertainty (for 1 d.o.f.) is given. No uncertainty on $R_\text{max}$ is reported, as our computation grid is too sparse to resolve it.}
    \tablewidth{\columnwidth}
    \tablehead{
        & \multicolumn{3}{c}{\textsc{Talys} - \textsc{Sibyll~2.3} }
    }
    \startdata
    $\gamma$
        & \multicolumn{3}{r}{$-0.80_{-0.23}^{+0.27}$}
    \\
    $R_\text{max}$ (GV)
        & \multicolumn{3}{r}{$(1.6 \pm 0.2) \cdot 10^{9}$}
    \\
    $m$                 
        & \multicolumn{3}{r}{$4.2_{-0.6}^{+0.4}$}
    \\
    $\delta_E$
        & \multicolumn{3}{r}{$0.14_{-0.03}^{+0.00}$}
    \\
    \hline
    $f_A (\%)$
        & H  & He & N 
        \\
        & $0.0_{-0.0}^{+42.6}$
        & $82.0_{-6.4}^{+3.8}$
        & $17.3_{-1.1}^{+1.0}$ 
        \\
        & Si & \multicolumn2c{Fe}
        \\
        & $ 0.6 \pm 0.1$
        & \multicolumn2c{$ (2.0 \pm 0.8) \cdot 10^{-2}$}
    \\
    \hline
    $I^{9}_A (\%)$
        & H  & He & N 
        \\
        & $0.0_{-0.0}^{+1.2}$
        & $9.8_{-2.9}^{+2.8}$
        & $69.2_{-1.6}^{+1.5}$ 
        \\
        & Si & Fe
        \\
        & $17.9_{-3.5}^{+3.2}$
        & $3.2_{-1.3}^{+1.2}$
    \\
    \hline
    $\chi^2$ / dof
        & \multicolumn3r{27.0 / 21}
    \\
    \enddata
\end{deluxetable}

Our ``baseline'' case is defined (a posteriori) by the combination of \textsc{Talys} as disintegration and \textsc{Sibyll~2.3} as air-shower model, motivated by its lowest $\chi^2$ out of realistic disintegration model choices. The other model combinations are discussed in \Sec\ref{sec:model_dependence}. 

The parameter space is shown in \Fig\ref{fig:paramspace_3D} and the best fit values in \Tab\ref{tab:bestfits3D}. We note that the $\chi^2/$dof is close to one, whereas it was close to three in the earlier 2D fit with fixed  energy scale and different disintegration and air-shower models; this means that we now actually have a good fit, due to the free source evolution and floating energy scale. The contour in the $\gamma - R_{\text{max}}$ plane is similar to the flat evolution case. Although the $\gamma \approx 1$ corresponding to Fermi acceleration with diffusive escape is within the 95\% contour, the preferred spectral indices result in flat or almost monochromatic spectra $\gamma < 1$. In contrast to the previous 2D case, a floating $\delta_E$ allows for somewhat softer spectral indices.

The $R_{\text{max}} - m$ plane exhibits a low rigidity cutoff for every choice of the source evolution within the 95\% CL. This is required by the composition data, in particular the $\sigma(X_{\mathrm{max}})$, that suggests a clear separation among the mass spectra. 
This result can be interpreted as a signature of the preference of the data for the maximum-rigidity scenario with respect to the photo-disintegration one. The discrimination among these scenarios is one of the science goals of AugerPrime \citep{Aab:2016vlz}, and what we found constitutes a stronger result with respect to the 2D fit.

The $\gamma - m$ parameter plane exhibits a clear anti-correlation, as already noticed for example in \citet{Unger:2015laa,Taylor:2015rla}. Positive source evolutions ($m>0$) result in a pile up from more distant sources, effectively softening the spectrum at Earth. This pile up is compensated by harder spectra at the source. Contrariwise, a high density of local sources ($m<0$) allows for spectral indices compatible with Fermi acceleration. The result clearly favors positive evolutions, covering star-forming objects, GRBs and Blazars. The very hard spectra found in this case are consistent with what was found for example in \citet{Taylor:2015rla}. The $3 \sigma$ contours leave room for negatively evolving sources such as TDEs \citep{Biehl:2017hnb}.

The spectrum and composition corresponding to the best fit of our baseline model are reported in \Fig\ref{fig:bestfit_3D}, while the corresponding injection spectra at the source (including the respective errors) are illustrated in \Fig\ref{fig:injection_bestfit}. The pile-up effect from higher redshifts is clearly visible: While the injection spectrum is very hard ($\gamma = -0.8$), the propagated spectra are softer and have a stronger overlap. The best fit for the proton component is $0$, and the proton component in the propagated spectrum comes only from propagation. However, the shaded range in \Fig\ref{fig:injection_bestfit} indicates the uncertainty in the normalization, which still allows for a significant proton fraction, as this component is barely contained in the fit range.

\newpage
\subsection{Model dependence of the UHECR fit}
\label{sec:model_dependence}

\begin{figure*}
    \centering
    \includegraphics[width=0.8\textwidth]{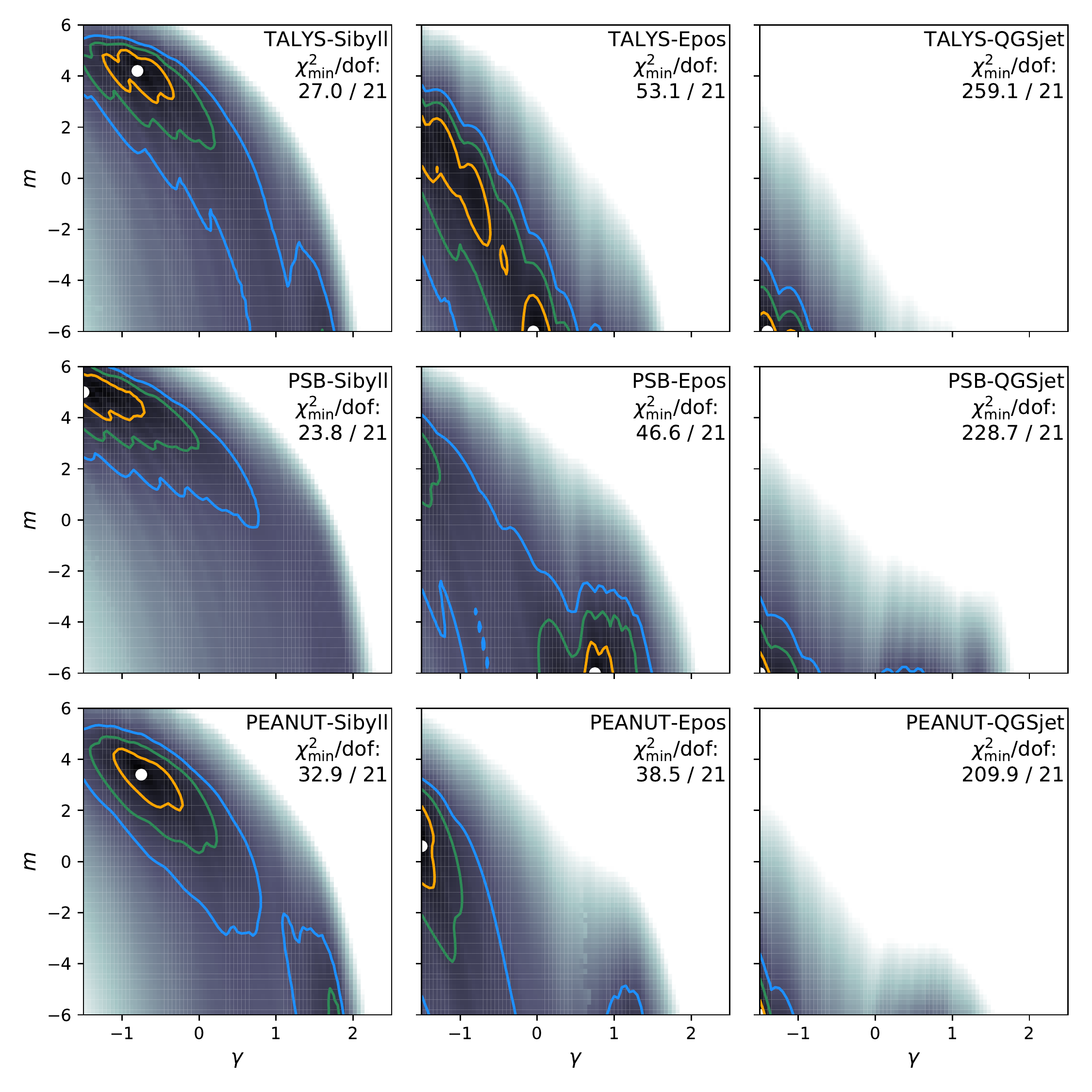}
    \caption{Parameter space in $\gamma$ and $m$ minimized over the third parameter $R_\text{max}$  for different combinations of disintegration and air-shower models. The color code and contours are defined as in \Fig~\ref{fig:paramspace_3D}.
    Rows from top to bottom: \textsc{Talys}, \textsc{PSB}, \textsc{Peanut}. Columns from left to right: \textsc{Sibyll 2.3}, \textsc{Epos-LHC}, \textsc{QGSjetII-04}. The corresponding best fit parameters can be found in \Tab\ref{tab:bestfits_all3D} (appendix).}
    \label{fig:paramspace_all_models}
\end{figure*}

We expand the discussion of the previous sections and study the influence of the propagation and air-shower models, by repeating the fit for permutations of the disintegration models \textsc{PSB}, \textsc{Talys}, \textsc{Peanut} and the air-shower models \textsc{Epos-LHC}, \textsc{Sibyll 2.3} and \textsc{QGSjetII-04}. The results are shown in \Fig\ref{fig:paramspace_all_models} for the projection to the $\gamma - m$ plane, and the corresponding best fit parameters are reported in \Tab\ref{tab:bestfits_all3D} (appendix).

Consistent with what was found in the CF, we cannot find reasonable fits for \textsc{QGSjetII-04} due to the model's broad $X_\text{max}$ distributions, in combination with a small $\langle X_\text{max} \rangle$, opposite to what is observed in data \citep{Auger2017xmax}. In all the other combinations we find satisfactory best fits with $\chi^2 / \text{dof} \approx 1.4-2.0$. Clearly, the shower model has a stronger impact on the fit contours than the disintegration model, as can be seen comparing the columns in \Fig\ref{fig:paramspace_all_models}. Interestingly, for the \textsc{PSB} model in combination with \textsc{Sibyll 2.3}, negative source evolutions are excluded at $3 \sigma$. This is an effect of the less efficient disintegration, as will be explained in the next section.

The anti-correlation between $m$ and $\gamma$ is found for all combination of disintegration and shower model (excluding \textsc{QGSjetII-04}). However, when exchanging \textsc{Sibyll 2.3} with \textsc{Epos-LHC}, the 3$\sigma$ contour in \Fig\ref{fig:paramspace_all_models} is shifted towards more local sources and/or more monochromatic spectra. 
The reason for this is that \textsc{Epos-LHC}, compared to \textsc{Sibyll 2.3}, predicts less shower-to-shower fluctuation decreasing the $\sigma(X_\text{xmax})$, while at the same time its $\langle X_\text{max} \rangle$ predicts a lighter composition of the measurements.
In combination this allows for less overlap of individual mass spectra.
Therefore local sources are favored for this model, reducing the impact of photo-disintegration, which would increase the mass overlap.
At the same time the maximal rigidity $R_{\text{max}}$ is more constrained for \textsc{Epos-LHC} than for \textsc{Sibyll 2.3} again decreasing the impact of photo-disintegration (this is not directly evident from \Fig\ref{fig:paramspace_all_models}).

The $\chi_\text{min}^2$/dof is slightly worse when using \textsc{Epos-LHC} ($\approx 2.0$) compared to \textsc{Sibyll~2.3} ($\approx 1.4$), mainly because the fit to the $\langle X_\text{max} \rangle$ is worse. It is however not strong enough to discriminate between these models, as the difference can be somewhat alleviated by allowing for shifts in $X_\text{max}$ within the systematic uncertainties. We did not include a proper treatment of these systematics.

Our results also show the limitations of what can be inferred from UHECR data alone. While the assumption of a generic rigidity-dependent source candidate describes the data sufficiently well, a strong degeneracy in the parameter space remains. Extending the range of the fit to lower energies could break this degeneracy, but would require assumptions about the extragalactic magnetic field and the transition to a (possibly) Galactic component below the ankle, which means that it would add more degrees of freedom to the model.

With new data from future experiments the situation is expected to improve. For example, with better information on the UHECR composition from the AugerPrime upgrade, the parameter space will likely be more constrained. A significant improvement of photo-disintegration and air-shower models would be needed as well; otherwise the ambiguity of the interpretation among different models will remain as indicated by our results.

\subsection{Injected composition}
\label{sec:injection_fractions}

\begin{figure*}
    \centering
    \includegraphics[width=.32\textwidth]{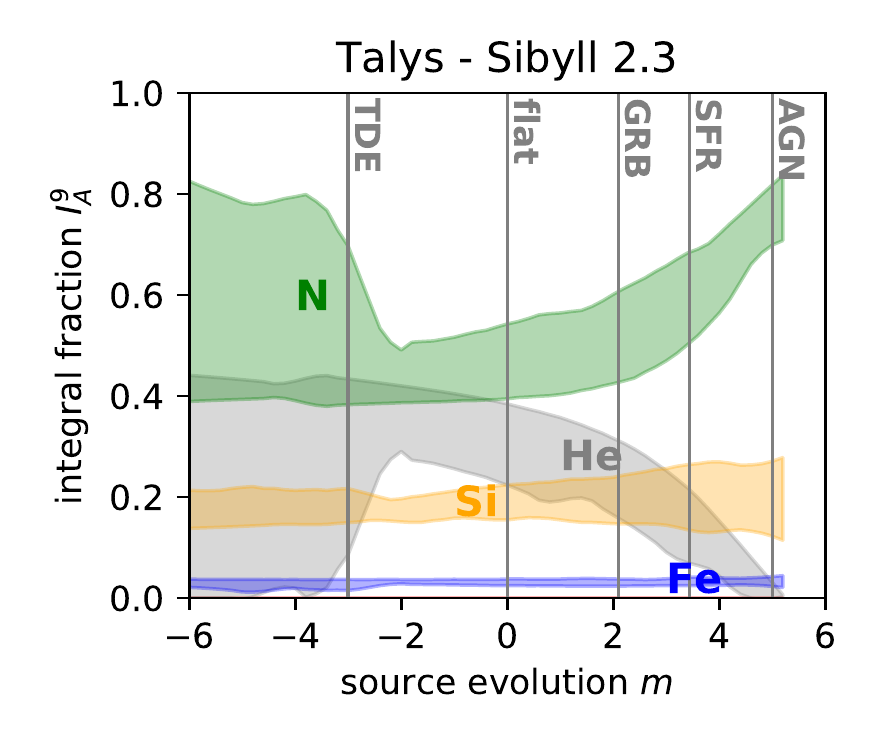}
    \includegraphics[width=.32\textwidth]{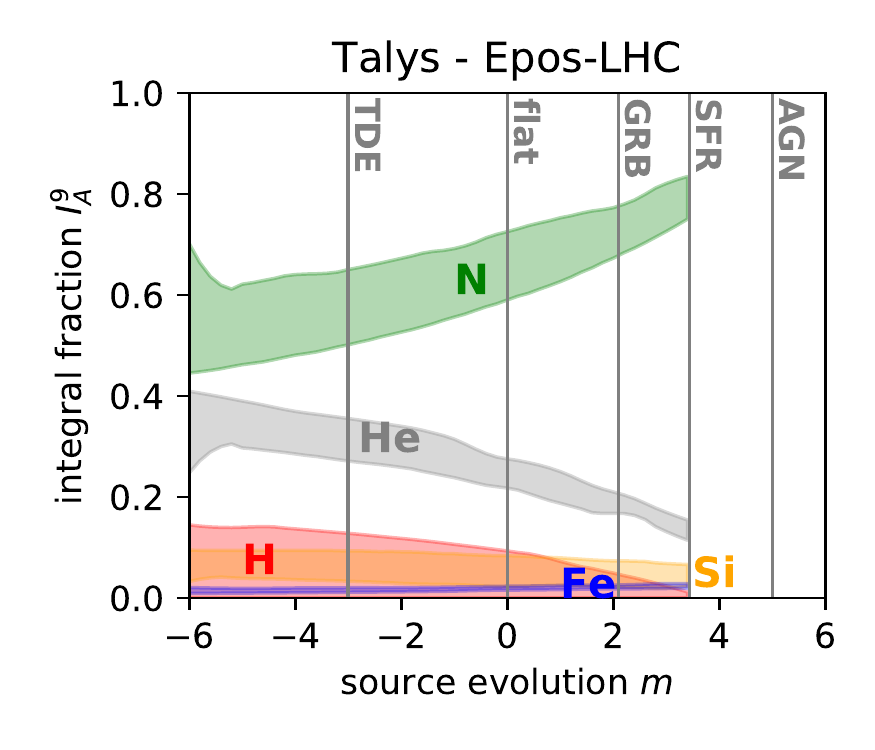}
    \includegraphics[width=.32\textwidth]{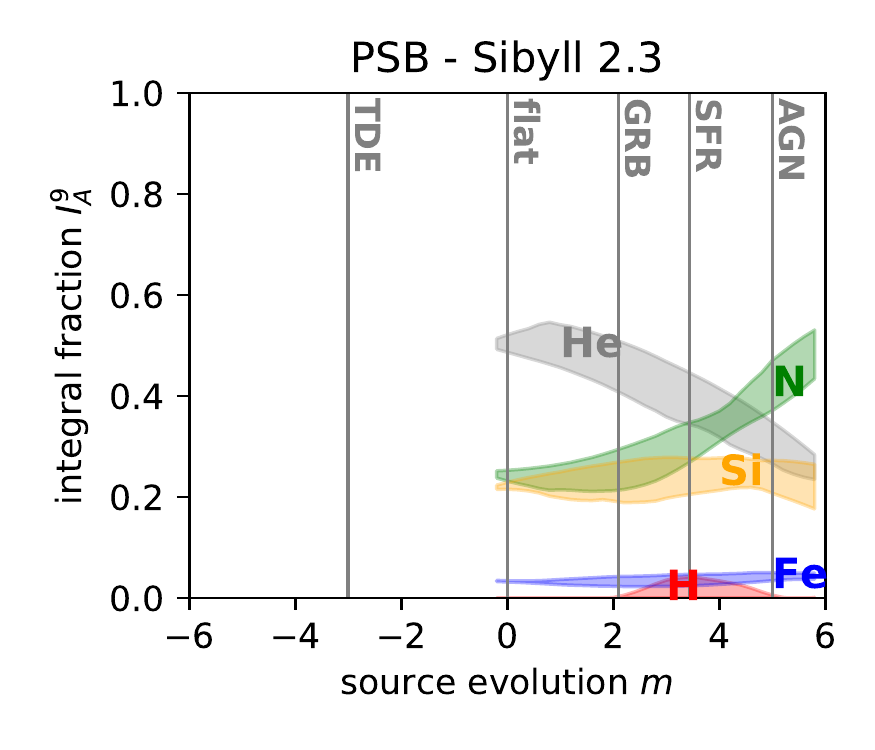}
    \caption{Ranges in the fraction allowed within $3 \sigma$ (for 2 d.o.f) as a function of the source evolution parameter. The fractions are defined by integrating the injection spectrum from $E_\text{min} = 10^9$ GeV, see \equ{integral_fractions}. Left: \textsc{Talys} - \textsc{Sibyll 2.3} (baseline model), Center: \textsc{Talys} - \textsc{Epos-LHC}, Right: \textsc{PSB} - \textsc{Sibyll 2.3}.}
    \label{fig:fractions_source_evol}
\end{figure*}

An interesting and reoccurring question is the range of mass compositions permitted by Auger data. While the composition at observation is fixed (within the uncertainty of air-shower models and data), it can have significantly different interpretations in terms of the composition ejected from the source.
Within the limitations of our model, we illustrate the ranges of the injected fractions $I^{9}_A$ within the $3 \sigma$ contours of our fit in \Fig\ref{fig:fractions_source_evol} as a function of the source evolution.
The figure shows the baseline case \textsc{Talys} - \textsc{Sibyll 2.3} as well as two additional panels changing the air-shower model to \textsc{Epos-LHC} and the disintegration model to \textsc{PSB}, respectively.

Comparing the fraction ranges for \textsc{Sibyll 2.3} (\Fig~\ref{fig:fractions_source_evol}, left) with respect to \textsc{Epos-LHC} (\Fig~\ref{fig:fractions_source_evol}, center) the most striking difference is in the silicon fraction, which is significantly higher for \textsc{Sibyll 2.3}, while in turn the nitrogen fraction is higher for \textsc{Epos-LHC}. This is mainly due to the heavier $\langle \ln A \rangle$ predicted by \textsc{Sibyll~2.3}. A significant proton fraction is only found in the case of \textsc{Epos-LHC}, owing to the slightly lower rigidity found for that model.
In both cases the nitrogen fraction increases at the cost of the helium fraction with higher source evolution. The higher disintegration for distant sources produces more helium during propagation, therefore requiring less helium injected at the source. 

For the same source evolutions, using \textsc{Sibyll~2.3} with respect to \textsc{Epos-LHC} leaves the mass fractions less constrained, as the combination of $\langle X_\text{max} \rangle$ and $\sigma(X_\text{max})$ predicted by \textsc{Sibyll~2.3} allows for a stronger superposition of different mass spectra.
In both cases the allowed mass fractions widen when going to negative source evolution.
This effect is directly connected to the propagation: for a larger concentration of distant sources the disintegration increases the spread of masses limiting the initial spread, while a larger concentration of local sources allows for a broader spread of isotopes already at the source.
This is an explicit demonstration that the $\sigma(X_{\mathrm{max}})$ reflects not only the spread of nuclear masses at the sources but also what happens during their propagation to Earth \citep{Abreu:2013env}.

The impact of the disintegration model is qualitatively different. As mentioned in \Sec\ref{sec:model_dependence}, negative source evolution is not contained in the $3 \sigma$ contours for the combination of \textsc{PSB} and \textsc{Sibyll~2.3}. This constrains the fraction ranges in \Fig\ref{fig:fractions_source_evol} (right panel) to positive source evolution.
The most relevant features of the disintegration model are the level of $\alpha$ emission and the number of open reaction channels that control how efficiently a nuclear cascade develops. For instance, the absence of $\alpha$ emission in \textsc{PSB}, is compensated by higher He fractions at the source, as noticed in \citet{Batista:2015mea,Aab:2016zth}. Due to the less efficient photo-disintegration in \textsc{PSB}, the necessary development of the nuclear cascade can be ensured only if the sources are distant enough (positive evolutions), leading to a rejection of local sources. This finding strengthens the need of using more refined models for photo-disintegration, since it demonstrates that the simple \textsc{PSB} model might bias the predictions for source evolution while overestimating the amount of helium at the source.

\Fig~\ref{fig:fractions_source_evol}, which describes the integral ejection fractions from the sources, can also be interpreted in terms of the physics of the sources. Especially the helium and proton fractions are indicative of the amount of disintegration required within the sources. While the isotopes must escape rather intact from the sources for strong evolutions, such as AGNs, weaker source evolutions seem to allow for higher helium and maybe even proton fractions -- which implies that the nuclei may partially disintegrate in the sources. While this gives a rough estimate, a rigid interpretation requires a more sophisticated source model.
For higher luminosity sources, that have a stronger disintegration chain, typically the rigidity-dependence of the maximal energy is not a valid assumption, see \eg\ \citet{Biehl:2017zlw,Rodrigues:2017fmu}.

A remarkable result is the non-zero iron fraction that we find throughout all model combinations. This is a result of the increased statistics at the cutoff of the updated Auger 2017 data set as discussed in \Sec\ref{sec:flat_evolution_fit}.

\section{Cosmogenic neutrino fluxes}
\label{sec:cosmo_neu}

\begin{figure}
    \centering
    \includegraphics[width=\columnwidth]{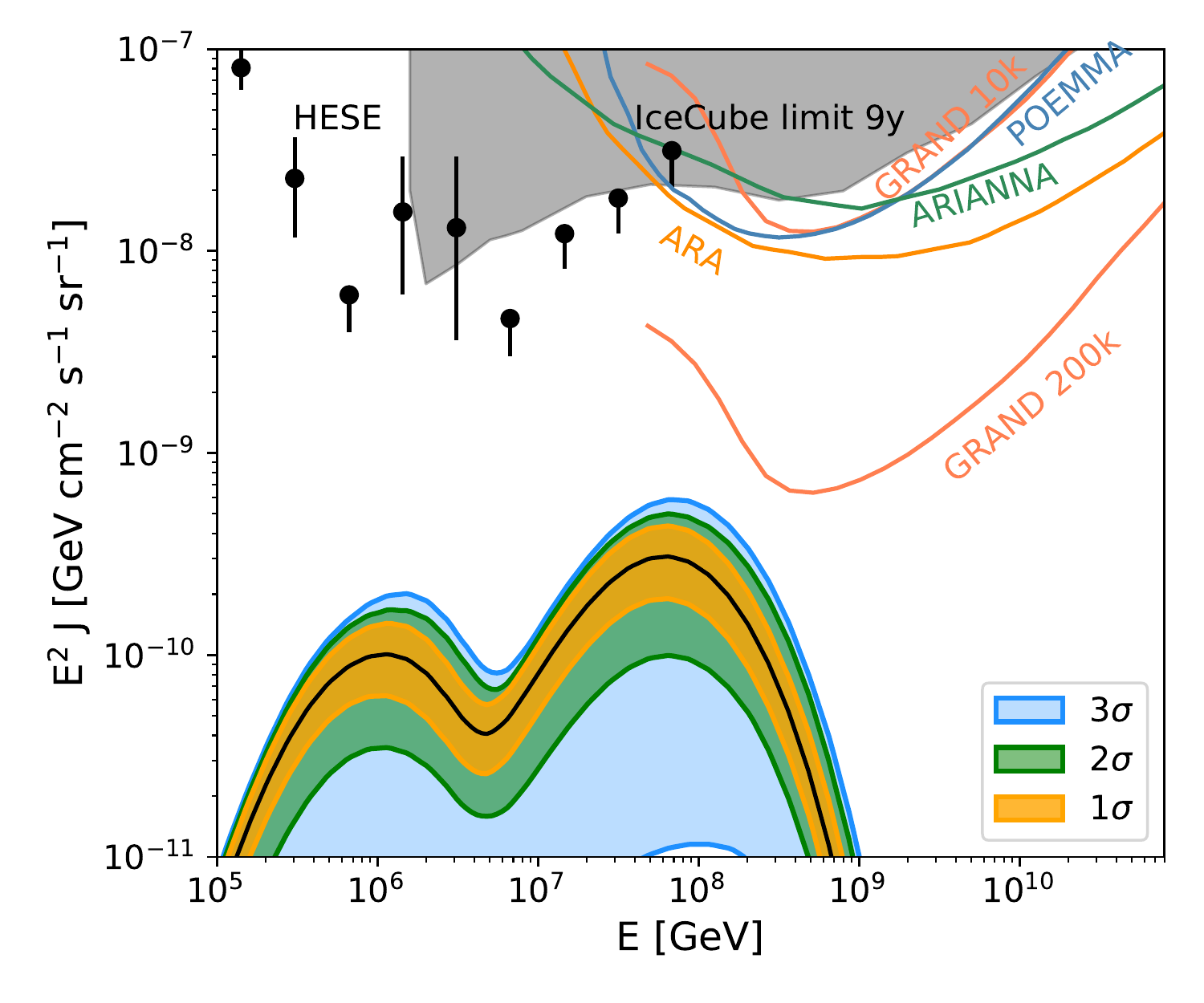}
    \caption{Allowed range of the neutrino flux (all flavors) from the 3D fit in \Fig\ref{fig:paramspace_3D} within the $1 \sigma, 2 \sigma, 3 \sigma$ contours (for 2 d.o.f.). The source evolution is in this case defined as $(1 + z)^m$ for a maximum redshift $z_\text{max} = 1$, where $m$ changes within the allowed fit regions. Estimated sensitivities for future radio neutrino detectors are shown for comparison: ARA~\citep{Allison:2015eky}, ARIANNA~\citep{Persichillithesis}, GRAND~\citep{Alvarez-Muniz:2018bhp} and POEMMA~\citep{KrizmanicUHECR2018}.}
    \label{fig:neutrinoflux_3D}
\end{figure}

\begin{figure*}
    \centering
    \includegraphics[width=\columnwidth]{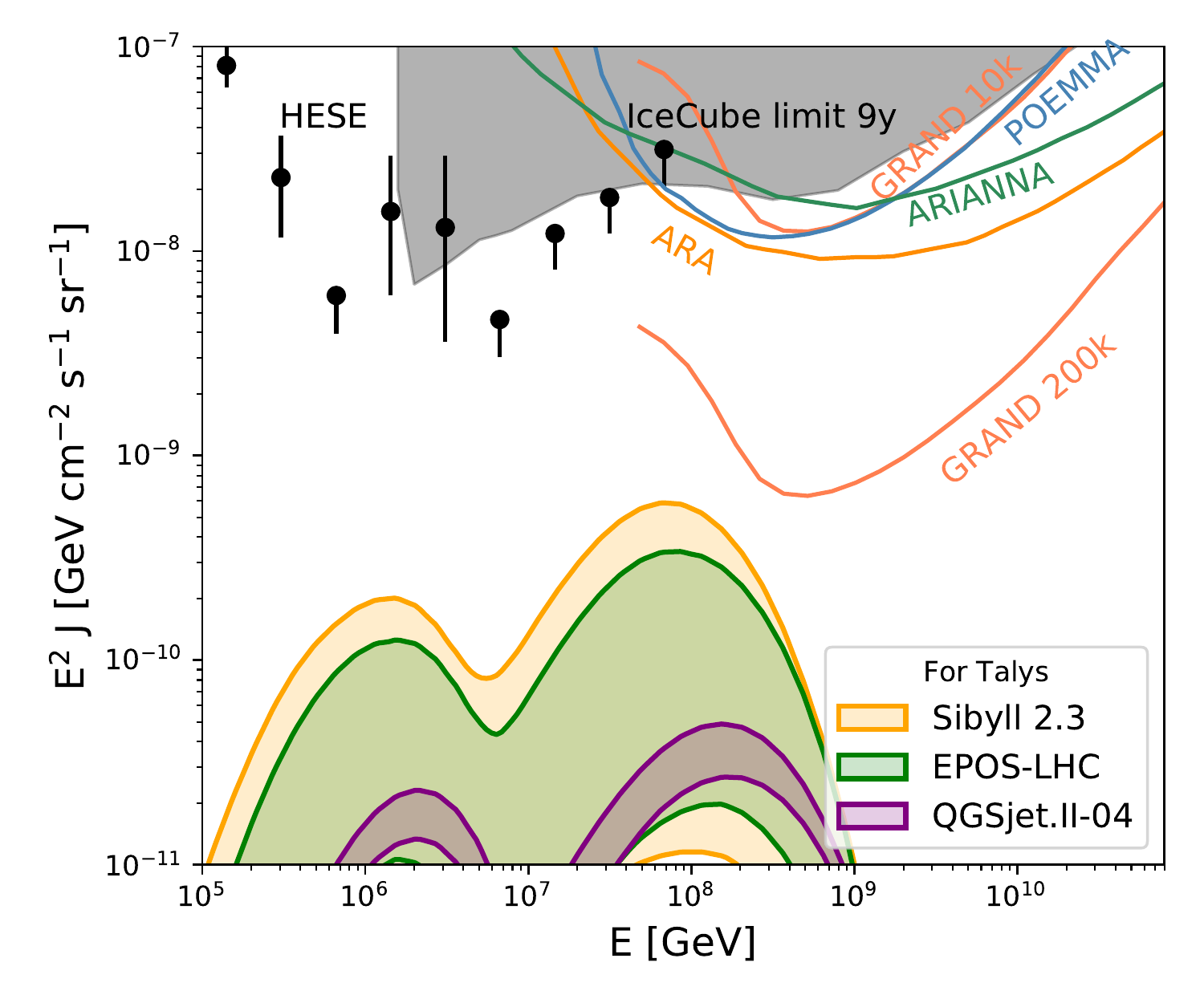}
    \includegraphics[width=\columnwidth]{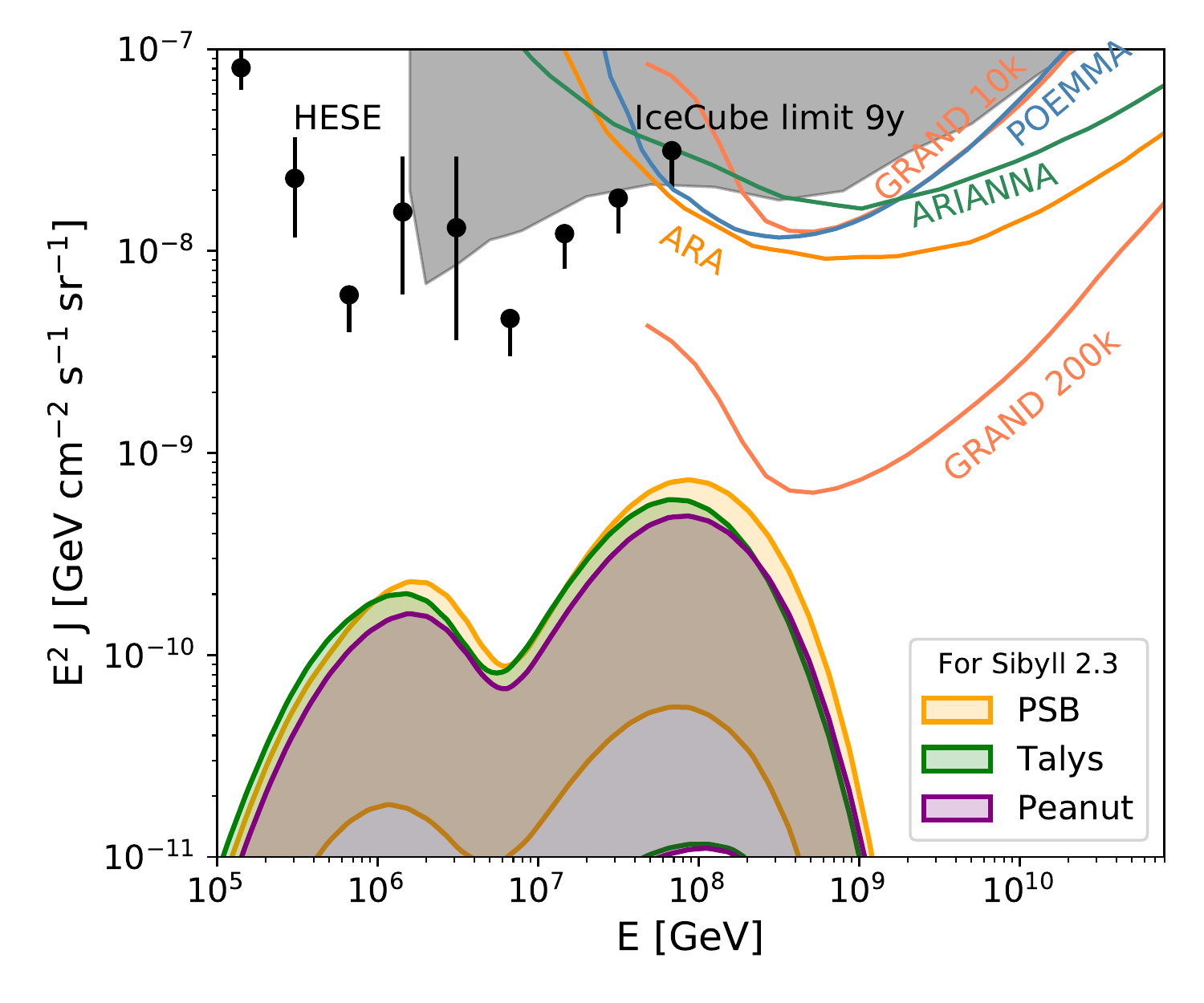}
    \caption{Allowed range for the neutrino flux (all flavors)  in the $3 \sigma$ region for different model assumptions. Left: The disintegration model is fixed to \textsc{Talys} and the ranges for different air-shower models are shown. Right: The shower model is fixed to \textsc{Sibyll 2.3} and the ranges for different disintegration models are shown.}
    \label{fig:neutrinoflux_model_dependence}
\end{figure*}

\begin{figure*}
    \centering
    \includegraphics[width=\columnwidth]{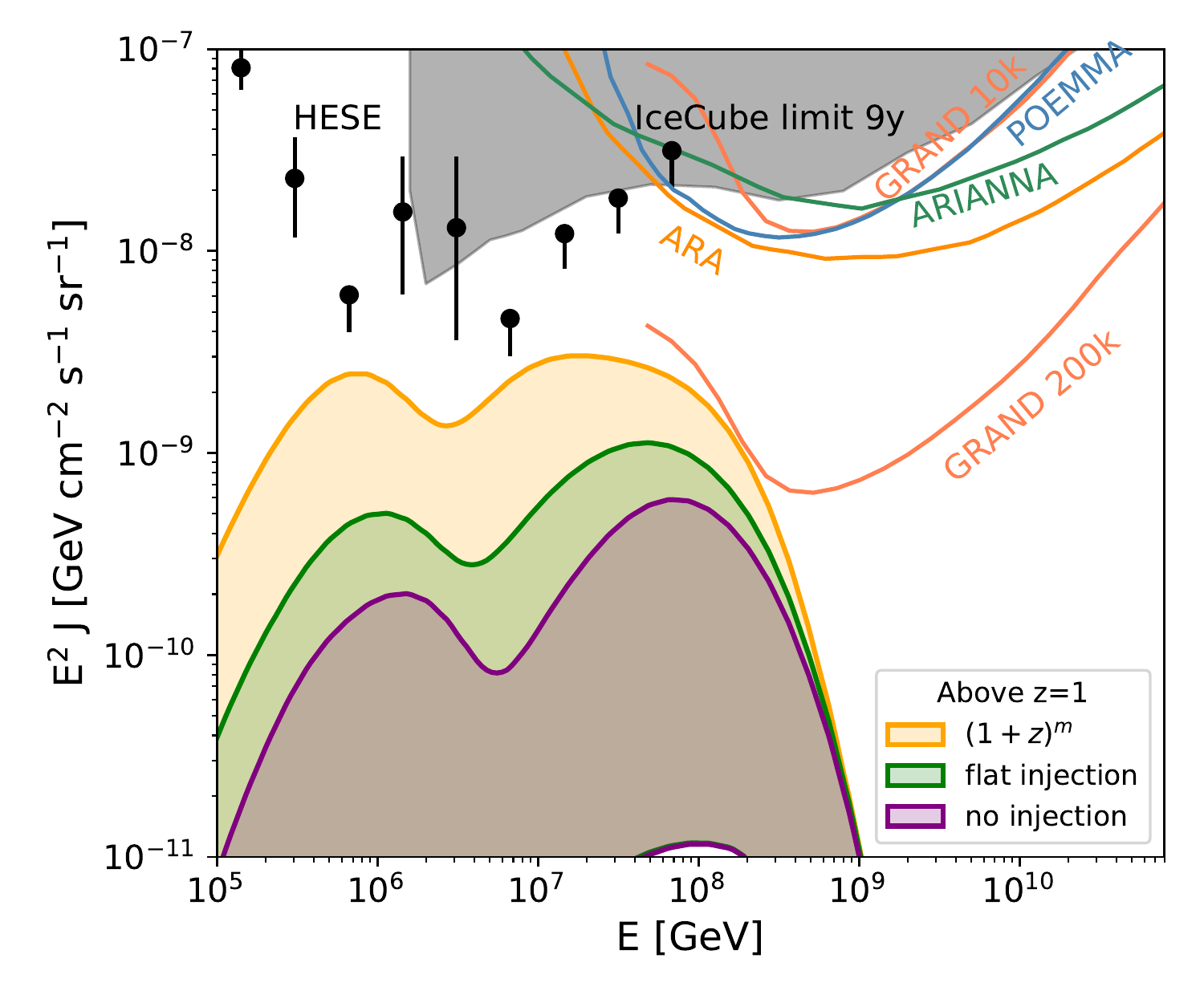}
    \includegraphics[width=\columnwidth]{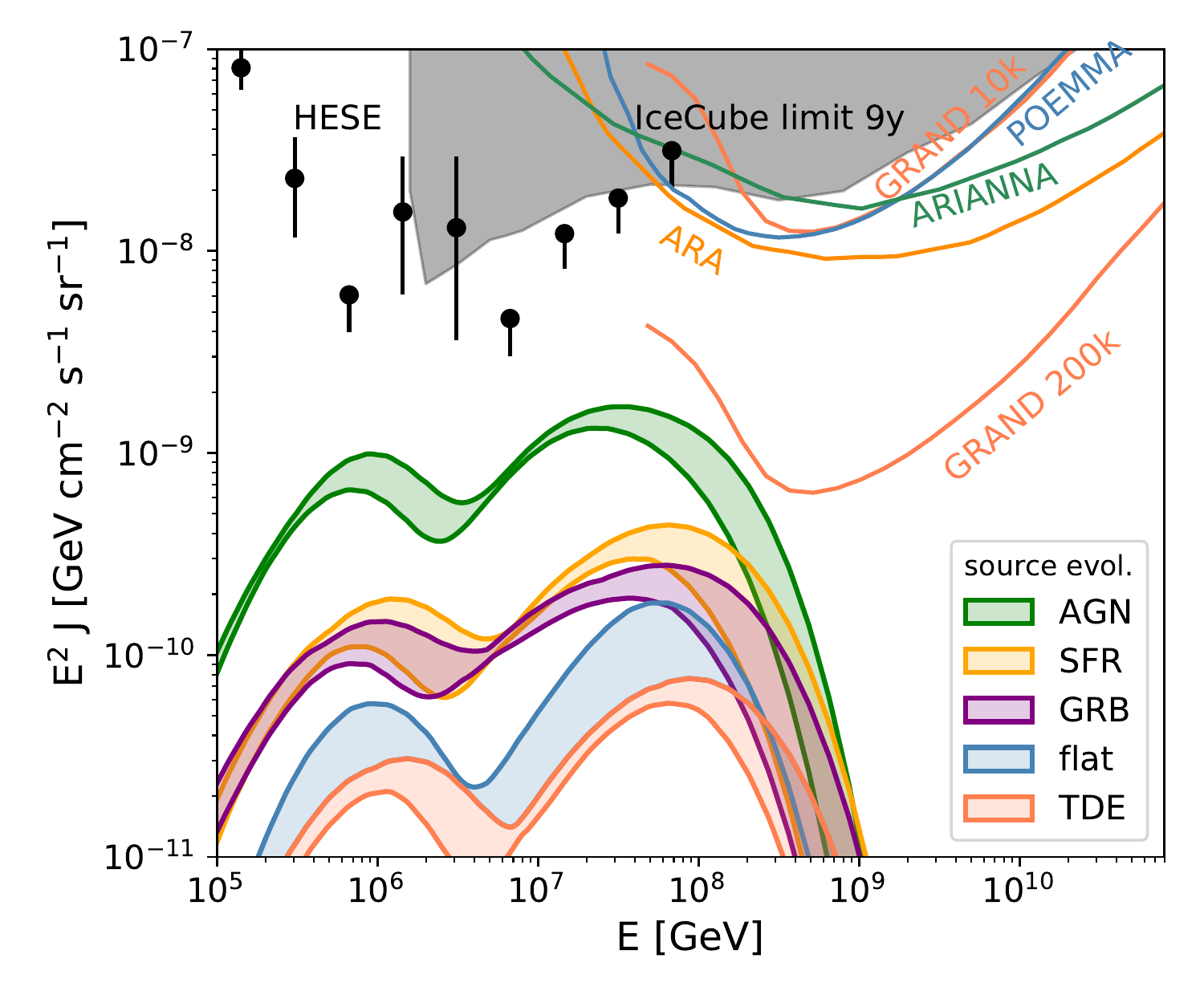}
    \caption{Allowed range for the neutrino flux (all flavors) in the $3 \sigma$ region for different source evolution. Left: The purple range corresponds to $z_\text{max} = 1$ (same as \Fig\ref{fig:neutrinoflux_3D}). For the other curves the source evolution is continued to $z_\text{max} = 3$ either by continuing as $(1+z)^m$ (yellow) or with a break to flat evolution at $z = 1$ (green). Right: The ranges are shown for the source evolution fixed to different source classes and for flat evolution.}
    \label{fig:neutrinoflux_3D_all_evo}
\end{figure*}

The source parameters inferred from the fit to UHECR data also lead to a prediction of the cosmogenic neutrino flux.
However, cosmogenic neutrino fluxes are significantly affected by the cosmic ray densities beyond a redshift of one, while UHECR fluxes are almost insensitive to such distant source populations. Therefore, it is impossible to estimate any confidence interval using a solely data-driven method. Under the assumption that the fit is sensitive up to a redshift of $z_\text{max} = 1$, we draw in \Fig\ref{fig:neutrinoflux_3D} the neutrino ranges corresponding to the 1, 2 and $3\sigma$ contours of the fit with the baseline model combination. Essentially, these flux levels can be regarded as constrained by present data. In contrast to the $1\sigma$ region, which is limited to positive source evolutions, the $3\sigma$ region is unconstrained towards negative redshifts (compare with \Fig\ref{fig:paramspace_3D}). Hence, if the sources are local, the expected cosmogenic fluxes are very low.

In the following we exclusively focus on the 3$\sigma$ contours. We study the robustness of our results against changes of the disintegration and the air-shower model in \Fig\ref{fig:neutrinoflux_model_dependence}. 
In the left (right) panel of \Fig\ref{fig:neutrinoflux_model_dependence} the cosmogenic neutrino flux is shown corresponding to the blue UHECR contours for the models in the top row (left column) of \Fig\ref{fig:paramspace_all_models}, respectively. The largest model dependence comes from the allowed range for the source evolution. The neutrino spectrum depends on the energy per nucleon, hence the composition dependence is weak. The variations between the disintegration models are small, resulting in a relatively robust upper bound. For \textsc{QGSjetII-04} the flux is small since positive evolutions are disfavored. For \textsc{PSB}, a sizable lower limit to the neutrino flux exists, since negative source evolution (local sources) is not allowed.

As the maximum rigidity is strongly constrained by the UHECR fit, the high-energy peak of the neutrino flux stays relatively robust and located at $\sim 10^8$~GeV. 
This is in agreement with \citet{AlvesBatista:2018zui}, where equally low fluxes were predicted.\footnote{Note that during completion of this work, an update to \citet{AlvesBatista:2018zui} was released, finding now the same low rigidities.}  
A small but relevant difference resides in the propagation code, since \citet{AlvesBatista:2018zui} assume a simplified redshift scaling of the CIB, whose effects in the neutrino fluxes are explained in \citet{AlvesBatista:2019rhs}. If we apply the same simplified scaling, the cosmogenic neutrino flux in our calculations increases by $50\%$. Other minor differences come from other details of the propagation code and the fitting procedure.
Differences to other works \citep{Romero-Wolf:2017xqe,Moller:2018isk,Das:2018ymz} come from their limiting assumptions about the source evolution, injected composition or the cutoff energy.

The most significant impact on the fluxes comes from the extrapolation to redshifts $z > 1$, which is unconstrained by UHECR data. For \Fig\ref{fig:neutrinoflux_3D_all_evo}, we adopt two approaches:
\begin{description}
    \item[(a)] (Left panel) an empirical method using a simple continuation of the $(1 + z)^m$ parameterization beyond $z=1$ up to $z_\text{max} = 3$. We also test a distribution with a break at $z=1$ and a flat ($m=0$) behavior beyond that.
    \item[(b)] (Right panel) discrete evolution functions of candidate source classes, where the parameter $m$ is not free; AGN \citep{Hasinger:2005sb,Stanev:2008un}, GRB \citep{2010MNRAS.406.1944W}, SFR \citep{Yuksel:2008cu} (including starburst galaxies), TDE \citep{Lunardini:2016xwi} and a flat evolution. In this case $z_\text{max} = 5$ is used, which is above the cutoff for all source evolutions used.
\end{description}

The most optimistic $(1+z)^m$ extrapolation results in fluxes that are one order of magnitude below the diffuse neutrino flux. It can be considered as the upper limit of what is expected in case of a single dominant UHECR source population with a rigidity-dependent energy cutoff. A flux at a similar level is found for AGN evolution.
In either scenario, the future radio-based instruments will neither be able to distinguish between source types (right panel) nor detect any significant cosmogenic neutrino signal. It is important to understand that the expected neutrino flux is (lower-) bounded only if the source evolution is fixed, motivated by a dominant source class. As long as the sources are not known or constrained, a ``minimal cosmogenic neutrino flux'' \citep{Ahlers:2012rz} is not meaningful. 

The low neutrino fluxes are partly related to our choice of generic source model, which leads to fits with low maximal rigidity. Other scenarios are possible in which a small fraction of the UHECR flux originates from proton accelerators that reach GZK energies \citep{vanVliet:2019nse}. These protons would copiously produce cosmogenic neutrinos off the denser CMB and peak at higher energies, while the majority of UHECRs would have a heavier mass composition, in line with current observations. These findings strongly support one of the science goals of the AugerPrime upgrade \citep{Aab:2016vlz}, in which additional hardware is deployed to determine the proton fraction among the observed UHECRs. This should be regarded as being of utter importance for the decisions regarding the next generation neutrino detectors. 
On the other hand this result leaves room for an unambiguous detection of very high energy neutrinos from the sources directly and it is unlikely that the cosmogenic flux will constitute a substantial background.

\section{Summary and conclusions}
\label{sec:conclusions}

In this work we have applied a  new numerical high-performance propagation code, \prince{}, 
to the updated spectrum and the composition data published by the Pierre Auger Observatory in 2017. We have included the source evolution $m$ as an additional free parameter. The savings in computation time have been used in favor of a detailed assessment of the main model dependencies, the nuclear disintegration in the propagation, and, the hadronic interactions in the air-shower development. For the emission from generic UHECR sources, we have retained the main assumption from the Combined Fit (CF) of a single dominant accelerator type. Our results, therefore, refer to an ``average'' or ``generic UHECR accelerator'' that emits nuclei at most as heavy as iron with a spectral cutoff at a maximal rigidity.

We have demonstrated that the reduced statistical error of the 2017 data set, in particular at the highest energy data points, favors for the first time a small but constrained iron fraction almost independent of the model variations. This implies a somewhat lower maximal rigidity.

The extension to three dimensions ($\gamma$, $R_\text{max}$ and $m$) confirms and strengthens the finding of a low $R_\text{max}$ independent of the source evolution.  
We find a clear indication of a correlation between the spectral index and source evolution:
rigidity-dependent source candidates must be local $m < 0$ with spectral indices compatible with those obtained in models with diffusive shock acceleration, or, distributed according to the star forming rate but with very hard, almost monochromatic, spectral indices. Source classes discussed in the literature, corresponding to such scenarios, are jetted Tidal Disruption Events~\citep{Zhang:2017hom,Guepin:2017abw,Biehl:2017hnb} and low luminosity GRBs~\citep{Zhang:2017moz,Boncioli:2018lrv}, or re-acceleration scenarios as those proposed for termination shocks in starburst and nearby radio galaxies \citep{Anchordoqui:2018vji,Eichmann:2017iyr,Winchen:2016koj}, respectively. While the inclusion of magnetic fields would soften the spectra at the source, the effect is probably not significant enough to draw an entirely different conclusion. It is challenging to reconcile this result with astrophysics, since a large number of alike sources with very similar $R_\text{max}$ and mass composition is needed to reproduce the observations. 

We have assessed the impact of model variations on the contours in the $\gamma$ -- $m$ plane for all combinations of the disintegration models \textsc{PSB}, \textsc{Peanut}, and \textsc{Talys} and the air-shower models \textsc{Epos-LHC}, \textsc{Sibyll 2.3}, and \textsc{QGSjetII-04}. The largest effect comes from changes in the air-shower modeling, which means that a better understanding of hadronic interactions would provide useful constraints. However, the 3$\sigma$ contours enclose the entire range of $m$, implying that there is no clear preference for a candidate source type. While the model variations lead to  unconstrained distributions of the source, their mass composition is limited, preferring a mixture of nitrogen and helium with an admixture of silicon depending on the level and efficiency of nuclear disintegration during the transport. We have shown that the use of simplified disintegration models prevents the possibility of investigating the whole parameter space including local sources. Other choices in the number or type of elements do not significantly affect the result.  

By using the contours that represent the compatibility with UHECR observations, we have studied the cosmogenic neutrino fluxes; compared to a purely theoretical prediction, this can be regarded as a postdiction from UHECR data. Because the allowed range in $m$ is unbounded, no meaningful lower bound can be derived for cosmogenic neutrinos since local sources cannot be excluded by the fit. On the other hand, we find that the upper bound is relatively robust under model variations. The fluxes are only constrained under fixed assumptions for the cosmic distribution of sources motivated by specific source classes. 

In all cases, the expected flux is small and peaks at energies around $10^8$ GeV making the detection by the proposed future radio-based detectors unlikely. On the other hand, this result means that if very high energy neutrinos from sources exist at energies beyond $10^8$ GeV, the expected background from diffuse cosmogenic neutrinos is expected to be small.
This conclusion applies if UHECRs are produced in one dominant type of accelerator with rigidity-dependent maximal energy cutoffs. If there are multiple types, for instance including a subset of proton rich sources, then the fluxes can look significantly different. Additional clues from high-precision composition measurements are highly valuable, which the AugerPrime upgrade is expected to deliver in a few years from now. 

In conclusion, the fit is relatively sensitive to the disintegration and, even more, the air-shower model, which still lead to a strong ambiguity in the intepretation of the data and therefore need future improvements. The predicted cosmogenic neutrino flux is relatively robust with respect to these models, and probably out of the reach of future experiments in all cases. A significant enhancement to the neutrino flux can come from redshifts beyond one, which cannot be constrained from  UHECR data alone.

{\bf Note:} During the completion of this work, \citet{AlvesBatista:2018zui} updated their manuscript; parts of our results have been made accessible to the authors of this paper, and are in agreement with the updated version (v2) of this manuscript.

\subsubsection*{Acknowledgments}
We thank A. van Vliet for useful feedback to the draft of this paper and T. Piran for inspiring discussions. We also thank the colleagues from the Pierre Auger Collaboration. This work has been supported by the European Research Council (ERC) under the European Union’s Horizon 2020 research and innovation programme (Grant No. 646623).

\bibliographystyle{aasjournal}
\bibliography{references}

\clearpage
\appendix
\section{Propagation Code - \prince{}}
\label{sec:prince_code}

For our study we have written an original computer code in order to have a framework in which systematic uncertainties such as cross sections and photon backgrounds can be efficiently varied. This appendix contains details about the numerical methods used to accelerate the computation of the UHECR transport equation.

The two popular public UHECR propagation codes (\textsc{CRPropa}~\citep{Batista:2016yrx} and \textsc{SimProp}~\citep{Aloisio:2017iyh}) use a Monte-Carlo approach. While these can effectively handle spectral properties by re-weighting samples, a rigorous treatment of certain systematics, such as photo-nuclear cross sections, requires a full computationally expensive re-sampling. On the other hand, an iterative numerical solution of the transport equation system, requires a constant computational time under the variation of any parameter. The trade-off is that the variation of spectral properties requires a full re-computation, as well.

Our code is called \prince{} (\textit{\textbf{Pr}opagation \textbf{i}ncluding \textbf{N}uclear \textbf{C}ascade \textbf{e}quations}). The main development goals were as follows:

\begin{itemize}
    \item A \textbf{time dependent UHECR transport equation solver} efficient enough to compute a single spectrum within seconds
    \item \textbf{Fast and easy variation of model input} such as cross section models and extragalactic photon backgrounds
    \item \textbf{Accessibility and modularity}, such that users can easily modify and extend specific parts of the code through interfaces. 
\end{itemize}

To achieve these goals, \prince{} is written in pure \textsc{Python} using vectorized expressions for the performance intensive parts, accelerating those using libraries like \textsc{Numpy} and \textsc{Scipy} \citep{Scipy}.
This vectorized approach also allows for the code to be implemented for massively parallel accelerators, such as  graphics processing units (GPUs), without much additional effort.

The Boltzmann transport equation for UHECRs is most conveniently solved in terms of the comoving density $Y_i(E_i, x, z) = N_i(E_i,x,z)/(1+z)^3$.
Assuming homogeneous and isotropic sources the diffusion terms vanish and the transport equation becomes independent of the spacial coordinate $x$ (Propagation theorem \citep{Aloisio:2004jda}). The coupled differential equation system for the particle species $i$ reads 
\begin{align}
\begin{split}
    \label{eq:transport_simple}
    \partial_t Y_i = 
    -\partial_E (b_\text{ad} Y_i) - \partial_E \left(b_{e^+ e^-}  Y_i \right)
    - \Gamma_i Y_i + \sum_j Q_{j \rightarrow i} (Y_j) + J_i ,
\end{split}
\end{align}
where we introduced the simplified notation ${Y_i = Y_i(E_i,z)}$, which can be transformed between time $t$ and redshift $z$ with the relation $dz = -dt (1 + z) H(z)$.
The first two terms describe the continuous energy losses due to adiabatic cooling ($H E$) and Bethe-Heitler pair-production ($b_{e^+ e^-}$).
$\Gamma_i$ is the rate of photo-nuclear interactions. The conversion of the particle species $j$ into $i$ is handled by the re-injection terms $Q_{j \rightarrow i}(Y_j)$. The decay terms for unstable particles can be treated implicitly, as described below.
The last term ($J_i$) describes the injection from sources. We will discuss the partial and ordinary differential parts separately in the following two sections.

\subsection{Photo-hadronic interactions: ODE}
\label{sec:photo-hadronic-ode}

Our approach to solve the ordinary differential equation (ODE) system that describes the conversion between particle species due to photo-nuclear interactions follows the method and the notation described in \citet{Boncioli:2016lkt,Biehl:2017zlw}. This new approach however greatly benefits from rewriting the same equations in terms of matrices.

The result a photo-nuclear interactions above a few MeV is the production of at least one or more final state particles, in which the projectile nucleus disintegrates. In the system of ODE the disintegration happens with the rate $\Gamma_i \equiv \Gamma_i(E_i)$ and the (re-)injection terms $Q_{j \rightarrow i} (Y_j,E_i)$ couple the equation systems of different particle species. The general form of the interaction rate on a target photon field is given by an integral over the photon energy $\varepsilon$ and the pitch angle $\theta$ in comoving frame
\begin{align}
\begin{split}
    \Gamma_i(E_i) = \int \text{d} \varepsilon \int_{-1}^{+1} &\frac{\text{d}\cos\theta}{2} (1 - \cos \theta)
    \, n_\gamma(\varepsilon, \cos\theta) \, \sigma_i(\varepsilon_r(\theta,E_i,\varepsilon)).
\end{split}
\end{align}
The $\sigma_i(\varepsilon_r)$ is the absorption (total) photo-nuclear interaction cross section as a function of the photon energy in the nuclear rest frame ${\varepsilon_r = (E_i \varepsilon) / m_i \cdot (1 - \cos\theta)}$. For isotropic photon fields the pitch-angle-averaged cross section $f(y)$ can be pre-computed and the interaction rate becomes
\begin{align}
\label{eq:interaction_rate}
\begin{split}
    \Gamma_i(E_i) &= \int \text{d} \varepsilon \, n_\gamma(\varepsilon) \, f_i(y(E_i,\varepsilon))\\
    f(y) &= \frac{1}{2 y} \int_{0}^{2 y} \text{d} \varepsilon_r \, \varepsilon_r \, \sigma_i(\varepsilon_r) ,
\end{split}
\end{align}
where $y \equiv (E_i \varepsilon)/m_i$ corresponds to the pitch-angle-averaged photon energy. The re-injection rate has a similar form, but expressed with the inclusive differential cross section ${\rm d}\sigma_{j \rightarrow i} / {\rm d} E_i (E_j,E_i,\varepsilon_r)$ and an additional integral over projectile densities $Y(E_j)$. The inclusive differential cross section can again be pitch-angle-averaged and expressed as a function of $y$. In analogy to \equ{interaction_rate} the re-injection rate reads
\begin{align}
\label{eq:reinjection_rate}   
\begin{split}
    Q_{j \rightarrow i}(Y_j, E_i) = \int_{E_i}^\infty &\text{d} E_j \, Y_j(E_j,z) \int \text{d} \varepsilon \, n_\gamma(\varepsilon)
    \, h_{j \rightarrow i}(E_i, E_j,y(E_j,\varepsilon))
\end{split}
\end{align}
with the kernel
\begin{align}
    \label{eq:h_func}   
    h_{j \rightarrow i}(E_i,E_j,y) = \frac{1}{2 y} \int_{0}^{2 y} \text{d} \varepsilon_r~ \varepsilon_r \deriv{\sigma_{j \rightarrow i}}{E_i}(E_j,E_i,\varepsilon_r) \, .
\end{align}
The decay of unstable particles is governed by a term $\partial_t Y_i = -\Gamma_{\text{dec},i}(E_i) Y_i$ with the decay rate $\Gamma_{\text{dec},i}(E_i) = (E_i/m_i~\tau_i)^{-1}$, where $\tau_i$ is the lifetime of an unstable particle or nucleus $i$ at rest. The re-injection terms for the decay products have a similar form to \equ{reinjection_rate}, but do not depend on the photon field. Hence the second integral can be omitted:
\begin{align}
\begin{split}
    \label{eq:particle_decays}
    Q_{\text{dec},j \rightarrow i}(Y_j, E_i) = \int_{E_i}^\infty \text{d} E_j \, &\Gamma_{\text{dec},j}(E_j) \,Y_j(E_j,z)
    \, \deriv{n_{j \rightarrow i}}{E_i}(E_j, E_i) \, .
\end{split}
\end{align}
The redistribution function $\text{d} n_{j \rightarrow i} / \text{d} E_i$ is in this case the inclusive energy distributions of the decay product $i$ in decays of $j$. To obtain inclusive distributions, all decay channels that contain $i$ are summed with their branching ratio as weight.

Most unstable particles that occur in UHECR propagation have a mean lifetime much smaller than the other relevant timescales. Hence, the decay can be regarded as an instant process at the production vertex. A decay chain via the intermediate meson or nucleus $u$, $j \rightarrow u \rightarrow i$, can be integrated out:
\begin{align}
\begin{split}
    \label{eq:decay_chain}
    \deriv{n_{j \rightarrow u \rightarrow i}}{E_i}(E_j, E_i) 
    = \int_{E_i}^{E_j} d E_u &\deriv{n_{j \rightarrow u}}{E_u}(E_j, E_u)
    \, \deriv{n_{u \rightarrow i}}{E_i}(E_u, E_i) \, .
\end{split}
\end{align}
For decay chains that proceed via multiple intermediate particles this formula is applied recursively. In practice, we substitute $\text{d} \sigma_{j \rightarrow u}/\text{d}{E_i}$ (production term for the unstable particle $u$) in \equ{reinjection_rate} with distributions of the decay products of $u$, $\text{d} \sigma_{j \rightarrow u \rightarrow X}/\text{d}{E_X}$ if $\tau_u < \tau_\text{thresh}$. For UHECR propagation we set $\tau_\text{thresh}$ to $\infty$, \ie all unstable particles decay immediately.

A special case arises for secondary nuclei. At high energies ($E_i \gg$ TeV), the impact of the internal nucleon motion can be neglected to a good approximation, resulting in the conservation of the boost of secondary fragments, \ie the energy per nucleon is conserved. The redistribution function then simplifies to
\begin{align}
    \label{eq:delta_func_per_nuclus}
    \deriv{\sigma_{j \rightarrow i}}{E_i}(E_j, E_i) \approx \sigma_j \, M_{j \rightarrow i} \, \delta \left(E_i - \frac{A_i}{A_j} E_j \right), 
\end{align}
where $M_{j \rightarrow i}$ is the averge multiplicity. For this case, it is convenient to express all equations in terms of energy per nucleon $E_i^A = E_i / A_i$. This leads to the simpler form of \equ{delta_func_per_nuclus}:

\begin{align}
    \label{eq:delta_func_per_nucleon}
    \deriv{\sigma_{j \rightarrow i}}{E_i^A}(E_j^A, E_i^A) \approx \sigma_j \, M_{j \rightarrow i} \, \delta (E_i^A - E_j^A), 
\end{align}
By computing the integral over $E_j$ for the reinjection rate in \equ{reinjection_rate} it simplifies to:

\begin{align}
    \begin{split}
        \label{eq:reinjection_boost_conserving}
        Q_{j \rightarrow i}(E_i^A) &= \frac{A_j}{A_i} \int \text{d} \varepsilon \, n_\gamma(\varepsilon) \, Y_j(E_i^A,z) \, g_{j \rightarrow i}(y) \\
        g_{j \rightarrow i}(y) &= \frac{1}{2 y} \int_{0}^{2 y} d \varepsilon_r \, \varepsilon_r \, M_{j \rightarrow i} \, \sigma(\varepsilon_r)
    \end{split}
\end{align}

For the discretization (see next section) it is convenient to formulate the equation system in $E_i^A$. This makes the treatment of the $\delta$-function in \equ{delta_func_per_nucleon} accurate as long as the same grid in $E_i^A$ is chosen for all nuclear particle species. We use the form \equ{reinjection_boost_conserving} for all nuclear species in the code. However for the sake of brevity we will not mention this explicitly in the following and only discuss the more general form \equ{reinjection_rate}.

\subsection{Discretization}
\label{sec:discretization}
For the numerical solution of the coupled ODE system \equ{transport_simple}, we introduce a discrete, logarithmic grid in energy:
\begin{align}
    E^k = E_0 \cdot 10^{k \cdot d_k} ,
\end{align}
where the grid constant $d_k$ can be adjusted independently for the particle and the photon grids to achieve the desired precision. Currently eight points per energy decade results in a good compromise between precision and computational speed. 
We use $k, l, m$ as upper indices for energy grid indices and $i, j$ as lower indices for particle species. All quantities are represented by their value at the interval centers. In some cases, such as for strongly peaked cross sections, it is necessary to compute precise averages over each interval instead of taking the central value.

On a grid we rewrite the interaction rate from \equ{interaction_rate} using step integrals as
\begin{align}
\begin{split}
\Gamma_i^k &= \Gamma_i(E_i^k) \\
    &= \sum_l \, \Delta\varepsilon^m \, f(E_i^k,\varepsilon^m) \, n(\varepsilon^m) \\
    &= \sum_l \, \Delta\varepsilon^m \, f_{i}^{km} \, n^m \\
    &= \mathcal{F}_i^k \cdot \vec{n} \, .
\end{split}
\end{align}
The factor $\Delta\varepsilon^m$ can be absorbed into either the kernel matrix $\mathcal{F}$ or the photon field vector $\vec{n}$. Here, we adopted the convention $\mathcal{F}_i^{km} = f_i^{km} \Delta \varepsilon^m$. The re-injection term from \equ{reinjection_rate} becomes
\begin{align}
\begin{split}
Q_{ji}^k &= Q_{j \rightarrow i}(Y_j, E_i^k) \\ 
    &= \sum_{l} \, \Delta E_j^l \, Y_j^l  \sum_m \, \Delta \varepsilon^m \, h(E_i^k,E_j^l,y^{km}) \, n(\varepsilon^m) \\
    &= \sum_{l} \, \Delta E_j^l \, Y_j^l  \sum_m \, \Delta \varepsilon^m \, h_{ij}^{klm} \, n^m \\
    &= \sum_{l} Y_j^l (\mathcal{H}_{i j}^{kl} \cdot \vec{n} ) \, ,
\end{split}
\end{align}
where the differential elements are absorbed into the kernel matrix $\mathcal{H}_{ij}^{klm} = h_{ijk}^{klm} \Delta E_j^l \Delta \varepsilon^m$. This allows us to write the coupled ODE system as a single matrix expression
\begin{align}
\label{eq:interaction_matrix}
\begin{split}
    \partial_t Y_i^k &= - \Gamma_i^k \, Y_i^k + \sum_j Q_{ij}^k + J_i^k \\
    &= - (\mathcal{F}_i^k \cdot \vec{n}) \, Y_i^k +  \sum_{j,l} \, (\mathcal{H} \cdot \vec{n} )_{i j}^{k l} \, Y_j^l+ J_i^k , \\
    \partial_t \vec{Y} &= \Phi \cdot \vec{Y} + \vec{J}.
\end{split}
\end{align}
The state vector $\vec{Y}$ contains all discretized particle spectra ordered by energy and particle mass, \ie:
\begin{align}
    \vec{Y} = (Y_{\nu_e}^0 \dots Y_{\nu_e}^K 
    \dots Y_{p}^0 \dots Y_{p}^K
    \dots Y_{\text{Fe}}^0 \dots Y_{\text{Fe}}^K
    )^T \, .
\end{align}
Although several symbols in the above equations appear tensor valued, we use an index translation scheme in the code that conveniently projects the equation system on a two-dimensional coefficient matrix $\Phi$, which is given by
\begin{align}
    \label{eq:interaction_matrix_structure}
    \Phi_{ij}^{kl} = \begin{cases}
        - (\mathcal{F}_i^k \cdot \vec{n}) + (\mathcal{H}_{ij}^{kl} \cdot \vec{n})
        & \text{if} \,\, i = j    \,\, \text{and} \,\, k = l \\
        \phantom{-} (\mathcal{H}_{ij}^{kl} \cdot \vec{n})
        & \text{if} \,\, i \neq j \,\, \text{or}  \,\, k \neq l
    \end{cases} .
\end{align}

Since each projectile produces only a few secondary particle species, the matrix $\Phi$ is sparse with only $\approx 2 \%$ of non-zero elements. The ordering of $\vec{Y}$ by energy and particle mass results in an upper-triangular shape of $\Phi$ and its sub-matrices, as long as there is no particle acceleration. The calculation of the derivative, a sparse-matrix vector dot-product, is significantly accelerated by using a sparse matrix storage format from a specialized library. The compact sparse row (CSR) format\footnote{An implementation of the CSR format is included in \textsc{Scipy}:\\ \url{https://docs.scipy.org/doc/scipy/reference/sparse.html}} stores a matrix $M$ as three vectors: A data vector $\vec{D}$ containing only the non-zero elements, a column index vector $\vec{C}$ holding the column indices for each element and a row pointer $\vec{R}$ pointing to the position of the first element of each row in $\vec{D}$ and $\vec{C}$. The end of each row is given by the next index in $\vec{R}$, an empty row is indicated by a repeated index in $\vec{R}$.

For example the matrix:
\begin{align}
    M = \begin{pmatrix}
        6 & 0 & 0 & 1 \\
        0 & 5 & 0 & 0 \\
        0 & 0 & 0 & 0 \\
        4 & 5 & 9 & 0  
    \end{pmatrix} ,
\end{align}
would be stored (with indexing starting at $0$) as:
\begin{align}
\begin{split}
    \vec{D} &= \begin{pmatrix}
        6 & 1 & 5 & 4 & 5 & 9 \\
    \end{pmatrix}\\
    \vec{C} &= \begin{pmatrix}
        0 & 3 & 1 & 0 & 1 & 2 \\
    \end{pmatrix}\\
    \vec{R} &= \begin{pmatrix}
        0 & 2 & 3 & 3 & 6 \\
    \end{pmatrix} .
\end{split}    
\end{align}

The format is to be read as following: The first two entries in $\vec{D}$ and $\vec{C}$ belong to the first row of $M$, as $R_1 = 2$ signals that the second row starts with the third entry. With $\vec{C}$ giving the column position, this means that $M_{0 0} = 6$ and $M_{0 3} = 1$. A repeated entry in $\vec{R}$ indicates an empty row, as for $R_2 = R_3 = 3$ in the example.
The vector $\vec{D}$ and $\vec{C}$ therefore always have a length equal to the number of non-zero elements, while $\vec{R}$ has a length equal to the number of rows plus one. The compact sparse column format (CSC) is defined analogously. The CSR format is especially effective multiplication with column vectors.

The vector $\vec{D}$ and $\vec{C}$ therefore always have a length equal to the number of non-zero elements, while $\vec{R}$ has a length equal to the number of rows plus one. The compact sparse column format (CSC) is defined analogously. The CSR format is especially effective for multiplication with column vectors.

In our approach the particle production channels and therefore the non-zero elements of $\Phi$ in \equ{interaction_matrix_structure} are fixed. Therefore the column index vector and row pointer only have to be found once. Instead of recomputing the whole sparsity structure, only the elements of the data vector in the sparse matrix format of $\Phi$ have to be replaced in every step, resulting in further computational speed gains. 

The computation of elements of $\Phi$ can be done in a single matrix expression if $\mathcal{F}$ and $\mathcal{H}$ are combined into a single cross section kernel $\mathcal{K}$. By ordering $\mathcal{K}$ according to the order of the $\vec{D}$ vector of $\Phi$, the elements of $\vec{D}$ can be modified in-place without additional memory allocations:

\begin{align}
    \label{eq:batch_matrix}
    \begin{split}
        &\phantom{=}\begin{pmatrix}
            \dots    &
            \Phi_{ii}^{kl} &
            \dots    &
            \Phi_{ij}^{kl} &
            \dots
        \end{pmatrix}^T
        = \mathcal{K} \cdot 
        \begin{pmatrix}
            n^{0}  \\
            \vdots \\
            n^{M}  \\
        \end{pmatrix}
        =\begin{pmatrix}
            \vdots  &  & \vdots \\        
            (-\mathcal{F}_{i}^{kl0} + \mathcal{H}_{ii}^{kl0}) & \dots & (-\mathcal{F}_{i}^{klM} + \mathcal{H}_{ii}^{klM})\\
            \vdots  &  & \vdots \\
            \mathcal{H}_{ij}^{kl0} & \dots & \mathcal{H}_{ij}^{klM} \\
            \vdots  &  & \vdots 
        \end{pmatrix}
        \cdot
        \begin{pmatrix}
            n^{0}  \\
            \vdots \\
            n^{M}  \\
        \end{pmatrix}
    \end{split}
    \end{align}
This arrangement allows for very fast computation of all coefficients of $\Phi$ and hence the handling of the time/redshift dependent ODE system becomes very efficient. The cross sections can be varied by scaling or replacing elements of the kernels in $\mathcal{K}$ between runs without additional initialization overhead.

\subsection{Adiabatic expansion and pair-production: PDE}
The partial differential part of the transport equation comes with two continuous loss terms:
\begin{align}
    \partial_t Y = -\partial_E (b_\text{ad} Y_i) - \partial_E \left(b_\text{pair}  Y_i \right)
\end{align}
with the loss terms $b \equiv d E / d t$.
The adiabatic losses due to cosmological expansion are described by
\begin{align}
    b_\text{ad} = - H(z) E.
\end{align}
Pair-production losses are implemented according to the continuous approximation by Blumenthal \citep{Blumenthal:1970nn}
\begin{align}
    b_{e^+ e^-} = &- \alpha r_0^2 Z^2 \, m_e^2
    \int_2^\infty d \xi \, n_\gamma \, \left( \frac{\xi m_e}{2 \gamma}, z \right) \frac{\Phi(\xi)}{\xi^2} .
\end{align}
We have already accelerated the numerical solution of photo-nuclear part of the equation system by solving the ODE in the sparse matrix form. The standard approach to include partial differential terms is to express the energy derivatives as finite differences, \eg\ second order central differences:
\begin{align}
    f^\prime (E_i^k) &= \frac{f(E_{i+1}^k) - f(E_{i-1}^k)}{2 \Delta E_i^k} + \mathcal{O}(({\Delta E_i^k})^2),
\end{align}
with $i$ indicating the energy grid index. For the entire energy grid, the finite differentiation operator can be written as a matrix:
\begin{align}
    D_i^{k ,k + 1} = \frac{1}{2 \Delta E_i^k} \qquad \qquad
    D_i^{k ,k - 1} = \frac{-1}{2 \Delta E_i^k}.
\end{align}
However this leads to an antisymmetric matrix, which has imaginary eigenvalues, in the case of second order differences. This leads to oscillations which can only be suppressed by using smaller step sizes. We find that it is equally accurate and more stable for our purpose to use forward biased differences, \eg\ in second order:
\begin{align}  
    f^\prime (E_i^k) &= \frac{-1 f(E_i^{k+2}) + 4 f(E_i^{k+1}) -3 f(E_i^{k})}{2 \Delta E_i^k} + \mathcal{O}(({\Delta E_i^k})^2) .
\end{align}
The code allows to adjust the order of finite differences to optimize for the given problem. Currently we use 6th order finite differences.
While this is probably more than necessary, we find that the impact on performance is small, as the computation time is dominated by the photo-hadronic part. For applications different from UHECR propagation we might however have to revisit this choice. If the order of the operator does not change, $D_i^{kl}$ can be included in the sparse interaction matrix $\Phi$ from \equ{interaction_matrix} that is solved as an ODE with methods described in the next section.

\subsection{Differential equation solver}
Using matrix formulation we have found an efficient scheme to recalculate the time derivative $\partial_t \vec{Y}(z)$. To solve the problem for $\vec{Y}(z)$ one has to choose an integration scheme in time $t$ (or for redshift $z$ by converting with $dz = -dt (1 + z) H(z)$).
For a system with light injection, the eigenvalues of the interaction matrix $\Phi$ are small enough such that we can use an explicit Euler scheme:
\begin{align}
    \vec{Y}(t + \Delta t) = \vec{Y}(t) + \Delta t \cdot \partial_t \vec{Y}(E,t)
\end{align}
For a proton system from redshift $z = 1$ with $dz = 10^{-3}$ the propagation can be solved within a few 100 ms.

For heavier mass nuclei the eigenvalues of $\Phi$ become very large and the system becomes stiff, requiring very small time/redshift steps for a stable explicit integration. In this case, we use an implicit integration scheme based on \textsc{scipy.integrate.ode.BDF} (\textit{Backward Differentiation}) solver, which adaptively adjusts the stepwidth and the order. A first order BDF sheme corresponds to an implicit Euler scheme:
\begin{align}
    \vec{Y}(t + \Delta t) - \vec{Y}(t) &= \Delta t \cdot \partial_t \vec{Y}(E,t + \Delta t) 
\end{align}
The arising implicit equation system is solved by Newton iteration to avoid inversion of the Jacobian in every step. More details are available in the \textsc{scipy} documentation\footnote{\url{https://docs.scipy.org/doc/scipy/reference/generated/scipy.integrate.BDF.html}}.

\begin{deluxetable*}{l|ccc|ccc|ccc}
    \centering
    \caption{\label{tab:bestfits_all3D} Best fit parameters for the 3D parameter scan with free source evolution for all nine model combinations, as described in \Sec\ref{sec:model_dependence}}
    \tablehead{
        & \multicolumn3c{\textsc{Talys} - \textsc{Sibyll~2.3}}
        & \multicolumn3c{\textsc{Talys} - \textsc{Epos-LHC}}
        & \multicolumn3c{\textsc{Talys} - \textsc{QGSjetII-04}}
    }
    \startdata
    $\gamma$
        & \multicolumn3{r|}{$-0.80_{-0.23}^{+0.27}$}
        & \multicolumn3{r|}{$-0.05_{-1.45}^{+0.10}$}
        & \multicolumn3r{$-1.40_{-0.10}^{+0.07}$}
    \\
    $R_\text{max}$ (GV)
        & \multicolumn3{r|}{$(1.6 \pm 0.2) \cdot 10^{9}$}
        & \multicolumn3{r|}{$2.5_{-0.9}^{+0.0} \cdot 10^{9}$}
        & \multicolumn3r{$1.8_{-0.0}^{+0.2} \cdot 10^{9}$}
    \\
    $m$
        & \multicolumn3{r|}{$4.2_{-0.6}^{+0.4}$}
        & \multicolumn3{r|}{$-6.0_{-0.0}^{+8.0}$}
        & \multicolumn3r{$-6.0_{-0.0}^{+0.2}$}
    \\
    $\delta_E$
        & \multicolumn3{r|}{$0.14_{-0.03}^{+0.00}$}
        & \multicolumn3{r|}{$0.11_{-0.01}^{+0.03}$}
        & \multicolumn3r{$0.14_{-0.01}^{+0.00}$}
    \\
    \hline
    $f_A (\%)$
        & H  & He & N 
        & H  & He & N 
        & H  & He & N
        \\
        & $0.0_{-0.0}^{+42.6}$
        & $82.0_{-6.4}^{+3.8}$
        & $17.3_{-1.1}^{+1.0}$
        & $0.0_{-0.0}^{+14.3}$
        & $90.0_{-0.4}^{+0.4}$
        & $9.8_{-0.4}^{+0.4}$
        & $82.2_{-1.5}^{+1.3}$
        & $17.3_{-0.9}^{+0.9}$
        & $0.5_{-0.0}^{+0.0}$
        \\
        & Si & Fe &
        & Si & Fe &
        & Si & Fe &
        \\
        & $0.6_{-0.1}^{+0.1}$
        & $0.0_{-0.0}^{+0.0}$
        &
        & $0.3_{-0.1}^{+0.1}$
        & $0.0_{-0.0}^{+0.0}$
        &
        & $0.0_{-0.0}^{+0.0}$
        & $0.0_{-0.0}^{+0.0}$
    \\
    \hline
    $I^{9}_A (\%)$
        & H & He & N
        & H & He & N
        & H & He & N
        \\
        & $0.0_{-0.0}^{+1.2}$
        & $9.8_{-2.9}^{+2.8}$
        & $69.2_{-1.6}^{+1.5}$
        & $0.0_{-0.0}^{+1.6}$
        & $38.1_{-1.0}^{+0.9}$
        & $54.2_{-1.1}^{+1.0}$
        & $12.3_{-1.1}^{+1.0}$
        & $27.6_{-1.3}^{+1.2}$
        & $58.3_{-0.5}^{+0.5}$
        \\
        & Si & Fe &
        & Si & Fe &
        & Si & Fe &
        \\
        & $17.9_{-3.5}^{+3.2}$
        & $3.2_{-1.3}^{+1.2}$
        &
        & $6.4_{-1.8}^{+1.8}$
        & $1.4_{-0.7}^{+0.7}$
        &
        & $0.0_{-0.0}^{+1.4}$
        & $1.8_{-0.3}^{+0.3}$
        &
    \\
    \hline
    $\chi^2$ / dof
        & \multicolumn3{r|}{27.0 / 21}
        & \multicolumn3{r|}{53.1 / 21}
        & \multicolumn3r{259.1 / 21}
    \\
    \hline \hline
    & \multicolumn3c{\textsc{PSB} - \textsc{Sibyll~2.3}}
    & \multicolumn3c{\textsc{PSB} - \textsc{Epos-LHC}}
    & \multicolumn3c{\textsc{PSB} - \textsc{QGSjetII-04}}
    \\
    \hline
    $\gamma$
        & \multicolumn3{r|}{$-1.50_{-0.00}^{+0.55}$}
        & \multicolumn3{r|}{$0.75_{-0.09}^{+0.12}$}
        & \multicolumn3r{$-1.50_{-0.00}^{+0.05}$}
    \\
    $R_\text{max}$ (GV)
        & \multicolumn3{r|}{$1.4_{-0.0}^{+0.5} \cdot 10^{9}$}
        & \multicolumn3{r|}{$3.5_{-0.4}^{+0.5} \cdot 10^{9}$}
        & \multicolumn3r{$1.8_{-0.2}^{+0.1} \cdot 10^{9}$}
    \\
    $m$
        & \multicolumn3{r|}{$5.0_{-0.6}^{+0.4}$}
        & \multicolumn3{r|}{$-6.0_{-0.0}^{+0.4}$}
        & \multicolumn3r{$-6.0_{-0.0}^{+0.2}$}
    \\
    $\delta_E$
        & \multicolumn3{r|}{$0.14_{-0.11}^{+0.00}$}
        & \multicolumn3{r|}{$0.14_{-0.03}^{+0.00}$}
        & \multicolumn3r{$0.14_{-0.02}^{+0.00}$}
    \\
    \hline
    $f_A (\%)$
        & H  & He & N 
        & H  & He & N 
        & H  & He & N
        \\
        & $0.0_{-0.0}^{+37.2}$
        & $98.5_{-0.1}^{+0.1}$
        & $1.4_{-0.2}^{+0.2}$
        & $0.0_{-0.0}^{+5.9}$
        & $87.8_{-0.2}^{+0.2}$
        & $11.1_{-0.6}^{+0.6}$
        & $83.7_{-0.8}^{+0.7}$
        & $16.1_{-0.4}^{+0.4}$
        & $0.2_{-0.0}^{+0.0}$
        \\
        & Si & Fe &
        & Si & Fe &
        & Si & Fe &
        \\
        &$0.1_{-0.0}^{+0.0}$
        &$0.0_{-0.0}^{+0.0}$
        &
        &$1.0_{-0.3}^{+0.3}$
        &$0.1_{-0.1}^{+0.1}$
        &
        &$0.0_{-0.0}^{+0.0}$
        &$0.0_{-0.0}^{+0.0}$
        &
    \\
    \hline
    $I^{9}_A (\%)$
        & H & He & N
        & H & He & N
        & H & He & N
        \\
        & $0.0_{-0.0}^{+1.8}$
        & $34.7_{-1.3}^{+1.3}$
        & $40.7_{-2.7}^{+2.5}$
        & $0.0_{-0.0}^{+1.5}$
        & $55.9_{-0.5}^{+0.5}$
        & $34.9_{-1.4}^{+1.4}$
        & $19.4_{-0.9}^{+0.9}$
        & $42.7_{-0.8}^{+0.8}$
        & $34.8_{-1.0}^{+1.0}$
        \\
        & Si & Fe &
        & Si & Fe &
        & Si & Fe &
        \\
        & $21.2_{-3.7}^{+3.4}$
        & $3.4_{-1.7}^{+1.7}$
        &
        & $7.6_{-2.1}^{+2.0}$
        & $1.5_{-1.0}^{+0.9}$
        &
        & $0.9_{-0.9}^{+1.6}$
        & $2.1_{-0.8}^{+0.8}$
        &
    \\
    \hline
    $\chi^2$ / dof
        & \multicolumn3{r|}{23.8 / 21}
        & \multicolumn3{r|}{46.6 / 21}
        & \multicolumn3r{228.8 / 21}
    \\
    \hline \hline
    & \multicolumn3c{\textsc{Peanut} - \textsc{Sibyll~2.3}}
    & \multicolumn3c{\textsc{Peanut} - \textsc{Epos-LHC}}
    & \multicolumn3c{\textsc{Peanut} - \textsc{QGSjetII-04}}
    \\
    \hline
    $\gamma$
        & \multicolumn3{r|}{$-0.75_{-0.21}^{+0.34}$}
        & \multicolumn3{r|}{$-1.50_{-0.00}^{+0.08}$}
        & \multicolumn3{r}{$-1.50_{-0.00}^{+0.03}$}
    \\
    $R_\text{max}$ (GV)
        & \multicolumn3{r|}{$1.8_{-0.1}^{+0.3} \cdot 10^{9}$}
        & \multicolumn3{r|}{$1.6_{-0.0}^{+0.2} \cdot 10^{9}$}
        & \multicolumn3{r}{$1.8_{-0.0}^{+0.2} \cdot 10^{9}$}
    \\
    $m$
        & \multicolumn3{r|}{$3.4_{-0.6}^{+0.6}$}
        & \multicolumn3{r|}{$0.6_{-0.8}^{+0.6}$}
        & \multicolumn3{r}{$-6.0_{-0.0}^{+0.2}$}
    \\
    $\delta_E$
        & \multicolumn3{r|}{$0.01_{-0.04}^{+0.03}$}
        & \multicolumn3{r|}{$0.14_{-0.01}^{+0.00}$}
        & \multicolumn3{r}{$0.14_{-0.00}^{+0.00}$}
    \\
    \hline
    $f_A (\%)$
        & H  & He & N 
        & H  & He & N 
        & H  & He & N
        \\
        & $0.0_{-0.0}^{+18.8}$
        & $93.8_{-0.5}^{+0.5}$
        & $5.7_{-0.5}^{+0.5}$
        & $62.3_{-8.3}^{+5.8}$
        & $37.1_{-1.3}^{+1.2}$
        & $0.7_{-0.0}^{+0.0}$
        & $84.7_{-0.9}^{+0.8}$
        & $15.1_{-0.5}^{+0.5}$
        & $0.2_{-0.0}^{+0.0}$
        \\
        & Si & Fe &
        & Si & Fe &
        & Si & Fe &
        \\
        & $0.4_{-0.1}^{+0.1}$
        & $0.0_{-0.0}^{+0.0}$
        &
        & $0.0_{-0.0}^{+0.0}$
        & $0.0_{-0.0}^{+0.0}$
        &
        & $0.0_{-0.0}^{+0.0}$
        & $0.0_{-0.0}^{+0.0}$
        &
    \\
    \hline
    $I^{9}_A (\%)$
        & H & He & N
        & H & He & N
        & H & He & N
        \\
        & $0.0_{-0.0}^{+0.9}$
        & $24.9_{-1.6}^{+1.6}$
        & $47.8_{-2.4}^{+2.2}$
        & $5.2_{-1.4}^{+1.4}$
        & $35.3_{-1.3}^{+1.2}$
        & $50.3_{-1.4}^{+1.3}$
        & $17.2_{-0.9}^{+0.9}$
        & $34.9_{-1.0}^{+1.0}$
        & $44.4_{-1.1}^{+1.0}$
        \\
        & Si & Fe &
        & Si & Fe &
        & Si & Fe &
        \\
        & $24.5_{-3.3}^{+3.0}$
        & $2.8_{-1.4}^{+1.3}$
        &
        & $8.0_{-2.3}^{+2.2}$
        & $1.2_{-0.9}^{+0.9}$
        &
        & $2.3_{-1.6}^{+1.5}$
        & $1.3_{-0.7}^{+0.6}$
        &
    \\
    \hline
    $\chi^2$ / dof
        & \multicolumn3{r|}{32.9 / 21}
        & \multicolumn3{r|}{38.5 / 21}
        & \multicolumn3r{209.9 / 21}
    \\
    \enddata
\end{deluxetable*}

\end{document}